\documentclass[a4paper]{aa}
\pdfoutput=1 
\usepackage{graphicx,color}
\usepackage{amsmath}
\usepackage{amssymb}
\usepackage{booktabs}
\usepackage[normalem]{ulem}
\usepackage[T1]{fontenc}
\usepackage{subcaption}
\usepackage{subfiles}
\usepackage[export]{adjustbox}
\usepackage{xcolor}
\usepackage{xspace}
\usepackage{euclid}
\usepackage{siunitx}
\usepackage[normalem]{ulem}
\usepackage{float}
\usepackage[utf8]{inputenc}
\usepackage{dblfloatfix} 
\usepackage[switch, modulo]{lineno}
\usepackage[pdfencoding=auto,psdextra]{hyperref}
\hypersetup{
    colorlinks=true,
    linkcolor=blue,
    filecolor=magenta,      
    urlcolor=blue,
    citecolor=blue
}
\newcommand{\Om}{\Omega_\mathrm{m}\xspace}

\newcommand{\Ob}{\Omega_\mathrm{b}\xspace}

\begin{document}
\title{Growth, geometry, and early-universe split of the matter density parameter $\Om$}
\titlerunning{Growth, geometry, and early-universe split of $\Om$}
\authorrunning{F. Keil, I. Tutusaus, \& A. Blanchard}
\author{Felicitas Keil$^{1}$\thanks{\email{felicitas.keil@utoulouse.fr}}, Isaac Tutusaus$^{2,3,1}$, and Alain Blanchard$^{1}$}
\institute{$^{1}$ Univ Toulouse, CNES, CNRS, IRAP, 14 Av. Édouard Belin, 31400 Toulouse, France \\
$^{2}$ Institute of Space Sciences (ICE, CSIC), Campus UAB, Carrer de Can Magrans, s/n, 08193 Barcelona, Spain \\
$^{3}$ Institut d'Estudis Espacials de Catalunya (IEEC), Edifici RDIT, Campus UPC, 08860 Castelldefels (Barcelona), Spain \\ }

\abstract{
While the $\Lambda$ cold dark matter ($\Lambda$CDM) model can successfully reproduce the measurements of many cosmological probes, some discrepancies have recently emerged. It is therefore necessary to test the standard cosmological model for consistency.
An important stress test is to separate the effect of different cosmological regimes on the parameter inference. We treat three regimes separately here: geometry, growth, and the early Universe. The geometric regime concerns the expansion and curvature history, the growth regime governs structure formation, and the early-universe regime affects physics prior to recombination. Previous analyses have performed the split between geometry and growth, whereas we also consider the influence of the early Universe separately.
We performed this split for the present-day matter density parameter $\Om$ using multiple cosmological observables. The data we used are galaxy clustering and weak-lensing statistics (3x2pt) from the Dark Energy Survey, cosmic microwave background data from \textit{Planck}, baryon acoustic oscillations from the Dark Energy Spectroscopic Instrument, type Ia supernovae samples from Pantheon+, and redshift-space distortions from a collection of galaxy surveys. For each of these probes, we introduced a phenomenological split into these three regimes.
The correlation between the geometric and the early regime for the matter density is strong, but that between the growth regime and the others is not. All regimes are compatible in the posterior distribution, but the difference between the geometry and the early regimes, $\Delta\Om^{\rm geo,early}$, has a 2$\sigma$ discrepancy with 0.
}

\maketitle

\section{Introduction}
The standard cosmological model, $\Lambda$CDM, assumes a cosmological constant $\Lambda$ that drives the accelerated expansion of the Universe and collisionless cold dark matter (CDM) that formed the large-scale structure. It is very successful in describing many different observations, such as galaxy clustering (GC), weak lensing (WL), the cosmic microwave background (CMB), and baryon acoustic oscillations (BAO) with only six free parameters. However, some discrepancies still remain. One example is the Hubble tension, which involves the discrepancy of the Hubble constant $H_0$ derived from CMB measurements by \Planck \citep{Planck:2018vyg, Planck:2019nip} versus local-universe measurements coming from type Ia supernovae (SNe Ia) in combination with Cepheids from Pantheon+ and the Supernovae and $H_0$ for the Equation of State of dark energy (SH0ES) data \citep{Brout:2022vxf, Riess:2021jrx}.

Another open question is dark energy, which is still not fundamentally explained. When modelled as a constant vacuum energy, it has to be fine-tuned in a way that cannot be explained by quantum field theory \citep{Weinberg:1988cp}. Recent BAO data from the Dark Energy Spectroscopic Instrument (DESI) DR2, combined with CMB and SNe Ia data, suggest a preference for non-constant dark energy at a significance of 2.8 - 4.2$\sigma$ depending on the SNe Ia dataset \citep{DESI:2025zgx}. In this case, the preferred equation of state using DESI DR2 shows a phantom crossing behaviour of dark energy. This violates the null-energy condition that requires that the dark energy density does not increase with the Universe expansion. Single scalar field models struggle to explain this \citep{Carroll:2003st}, but more complex models \citep{Kunz:2006wc} including some modified gravity models \citep{Koussour:2023ulc} would be able to accommodate this behaviour.

We tested the consistency of the $\Lambda$CDM model by comparing different sources of cosmological information. This has previously been proposed as a consistency test (see e.g. \cite{Zhang:2003ii, Ishak:2005zs, Abate:2008au, Ruiz:2014hma, Bernal:2015zom}, or \cite{Andrade:2021njl}). Furthermore, it has been applied to dark energy \citep{Perenon:2022fgw} and the neutrino mass \citep{Bertolez-Martinez:2024wez}. Cosmological observables contain information about geometry, structure formation, and physics in the early Universe. These regimes can be separated phenomenologically. The first regime concerns the geometric expansion of the Universe, which is related to cosmological distances, such as the comoving distance or the Hubble distance. The second regime is based on the structure growth of the Universe that arises from gravitational instabilities and does not account for the homogeneous expansion. Finally, the third regime is related to the physics of the early Universe before recombination.

We here focus on the present-day matter density and split the parameter $\Om$ into these three different regimes. The matter density is a good choice for this split because it is constrained by multiple cosmological surveys with a comparatively small uncertainty. Additionally, it affects cosmological observables in multiple different ways, which we used to split our likelihood into the three regimes. This parameter is also particularly interesting for a split because of the proposed so-called $\Om$ tension. This proposed tension mainly concerns the discrepancy between the matter density parameter in $\Lambda$CDM inferred from DESI BAO data \citep{DESI:2025zgx}, on the one hand, and the value inferred from CMB surveys, such as \textit{Planck} \citep{Planck:2018vyg}, and SNe Ia surveys, such as Pantheon+ \citep{Brout:2022vxf}, on the other hand. DESI DR2 finds $\Om = 0.2975 \pm 0.0086$, which differs at 2$\sigma$ from \textit{Planck}, which provides $\Om = 0.3153 \pm 0.0073$. The BAO dataset is also discrepant by 1.7$\sigma$ from Pantheon+, which lies at $\Om = 0.334 \pm 0.018$.
\cite{Colgain:2024mtg} showed that at the same effective redshift of $z\approx0.3$, the DES SNe Ia sample, DES-SN5YR \citep{DES:2024jxu}, and the DESI bright galaxy survey differ by $3.4\sigma$ in $\Om$. \cite{Colgain:2022rxy} further proposed that the detection of dynamical dark energy might be due to a matter density parameter that evolves with redshift. Here, we investigated possible discrepancies of $\Om$ at the same and at different redshifts.
 
A growth-geometry split was previously applied to the stage III galaxy surveys Dark Energy Survey \citep[DES,\footnote{\url{https://www.darkenergysurvey.org}}][]{PhysRevD.105.023520} by \cite{DES:2020iqt} and \cite{Zhong:2023how} and to the Kilo Degree Survey \citep[KiDS,\footnote{\url{https://kids.strw.leidenuniv.nl}}][]{KiDS:2020suj} by \cite{Ruiz-Zapatero:2021rzl}. In the latter, the authors split five cosmological parameters in $\Lambda$CDM, namely the reduced CDM density, $\omega_{\rm cdm}$, the rescaled amplitude of matter fluctuations, $S_8$, the reduced baryon density, $\omega_{\rm b}$, the spectral index, $n_{\rm s}$, and the reduced Hubble constant, $h$. They included BAO information in the geometric regime since it concerns cosmological distances. Thus, the sound horizon was based on the geometric matter density parameter. The authors also included redshift-space distortions (RSD) from the Baryon Oscillation Spectroscopic Survey (BOSS) DR12 \citep{BOSS:2016wmc} in the growth regime. From \textit{Planck} \citep{Planck:2018vyg}, the authors used CMB constraints on the scalar power spectrum amplitude $A_{\rm s}$ and on $n_{\rm s}$ as priors in the growth regime. These were classified as growth-related parameters since they capture primordial scalar fluctuations. The distance to the last scattering surface $\theta^*$ was considered geometric because of its relation to the background cosmology. The authors found that the two regimes were consistent. However, the splitting of all cosmological parameters led to comparatively large uncertainties.

For the DES survey, a growth-geometry split was first performed for DES-Y1 in \cite{DES:2020iqt}, splitting the parameter $\Om$ within $\Lambda$CDM. The authors also split the dark energy equation-of-state parameter, $w$, when they worked with the $w$CDM model. In addition to the joint analysis of GC, WL, and their cross-correlation (known as 3x2pt), the authors considered $\textit{Planck}$ CMB data marginalised over growth parameters based on \cite{Planck:2015bue}, SNe Ia from DES-SN3YR \citep{DES:2018ioi}, and BAO and RSD data from BOSS DR12 \citep{BOSS:2016wmc}. External data were treated as geometric, with the exception of RSD. The authors reported that geometry is much better constrained than growth. They used the Bayesian suspiciousness \citep{PhysRevD.100.043504} to quantify tensions and find discrepancies of 1.0 equivalent $\sigma$ for the split $\Om$ versus $\Lambda$CDM and 1.4 equivalent $\sigma$ for split $\Om$ and $w$ compared to $w$CDM.

\cite{Zhong:2023how} repeated this split for DES-Y3 data and slightly different external data, in particular, $\textit{Planck}$ CMB data with some scale cuts, SNe Ia from Pantheon \citep{Pan-STARRS1:2017jku}, BAO data from the Sloan Digital Sky Survey (SDSS) DR7 \citep{Ross:2014qpa} and 6dfGS \citep{Beutler:2011hx} surveys, and a Big Bang nucleosynthesis (BBN) prior \citep{Cooke:2016rky}. The authors found no sign of a tension between the growth and geometry regimes when they used all data. When they analysed DES 2x2pt data (the combination of GC and galaxy-galaxy lensing (GGL)) with the external datasets, they found a tension that was attributed to systematic effects. This approach was also used to  forecast the growth-geometry split for the  Legacy Survey of Space and Time \citep[LSST;][]{LSST:2008ijt} by \cite{Zhong:2024xuk}.

In addition to the growth and geometry regimes, we introduce a third and separate early-universe regime. This allows us to identify differences between the information from physics before recombination, the geometry of the Universe, and the growth of structures. In previous analyses, the early-universe physics were implicitly treated as geometry (in DES) or growth (in KiDS). Similarly to the DES case, we focused on splitting one parameter, $\Om$, to obtain better constraints and identify any possible discrepancy on the matter content from different regimes. 

The paper is organised as follows: in Sect.\,\ref{sec:methodology} we present the method and how we modelled each observable accounting for three regimes of the matter density. In Sect.\,\ref{sec:data} we describe the different datasets we considered in the analysis. We present our likelihood sampling in Sect.\,\ref{sec:likelihood} and show the results of the analysis in Sect.\,\ref{sec:results} 
before we conclude in Sect.\,\ref{sec:conclusions}.

\section{Method}\label{sec:methodology}
We split each of the probes that we used into different regimes and describe our ansatz here. SNe Ia data are an exception because they are only sensitive to the geometry regime. 

In the following, we describe this split in detail. Since this is a phenomenological approach, in some cases it is debated which regime should apply. This is evidenced by the fact that different approaches have been taken in the literature (see e.g. \citet{DES:2020iqt, Zhong:2023how, Ruiz-Zapatero:2021rzl}). Even with this caveat, this is still a null test that needs to be validated if $\Lambda$CDM is to be correct.

\subsection{Geometry, growth, and the early Universe}
The goal of this analysis is to split the effect of geometry, growth, and early-universe physics on the matter density parameter $\Om$. We receive information about these three regimes from cosmological probes, such as GC, WL, the CMB, BAO, SNe Ia, or RSD. Each of these probes is sensitive to one, two, or all three regimes in different ways, as we describe in Sect.\,\ref{observables}. 

We focused on the matter density parameter, $\Om$, because it is well constrained by current cosmological surveys and affects the observables we used in different ways that correspond to the three regimes. Considering additional parameters in the split can, in principle, provide useful complementary information, but it comes at the expense of larger uncertainties. We defer a dedicated study with additional split parameters to future work.

\subsection{Observables}
\label{observables}
\subsubsection{Photometric galaxy clustering and weak lensing (3x2pt)}
\label{photo-recipe}

For the joint analysis of photometric GC and WL, we followed an approach similar to that presented in the DES growth-geometry split\,\citep{DES:2020iqt, Zhong:2023how}, but with some modifications that we describe in the following.
One of the key ingredients in the 3x2pt analysis is the linear matter power spectrum, which depends on scale, $k$, and redshift, $z$. It is proportional to the square of the growth rate $D(z)$, \citep[see e.g. ][]{dodelson_modern_2003},
\begin{equation}
    P_{\rm lin}(k, z)=2\pi^2\delta_{\rm H}^2\frac{k^{n_{\rm s}}}{H_0^{n_{\rm s}+3}}T^2(k) \left(\frac{D(z)}{D(0)}\right)^2\,.
    \label{pk-lin-d-squared}
\end{equation}

Here, $\delta_{\rm H}$ is the scalar perturbation amplitude at horizon crossing, and $T(k)$ is the transfer function. The $H_0$ parameter stands for the Hubble constant. The growth factor $G(z)$ quantifies matter fluctuations, $\delta_{\rm m}$, that evolve with redshift, normalised by the initial matter fluctuations taken at an initial time around $z_{\rm ini}=1000$,

\begin{equation}
    G(z):=(1+z)D(z)=(1+z)\frac{\delta_{\rm m}(z)}{\delta_{\rm m}(z_{\rm ini})}\,.
\end{equation}

This quantity can be calculated using a second-order ordinary differential equation that depends on $\Om(\ln a)$ and $H(\ln a)$, with $a$ being the scale factor,
\begin{equation}
    G''+\left(4+\frac{H'}{H}\right)G'+\left(3+\frac{H'}{H}-\frac{3}{2}\Om\right)G=0\,,
    \label{growth-ode}
\end{equation}
where the prime denotes the derivative with respect to $\ln (a)$. This assumes that the growth is scale-independent. If there are massive neutrinos, this is slightly violated \citep[see, e.g.,][]{Hernandez:2016xci}. Furthermore, some modified gravity models, such as Dvali-Gabadzade-Porrati brane-induced gravity \citep{Dvali:2000hr}, predict a scale-dependent growth rate. For this consistency test, however, we relied on this approximation and neglected any possible scale dependence.

By default, $\Om^{\rm growth}$ is used to calculated the growth factor. We also describe below how this equation is used to rescale the power spectrum. When this formula is derived, the factors of $H(\ln a)$ and $H'(\ln a)$ come into play at multiple steps, that is, when the expressions are derived and when variables are substituted and rewritten. We omit the full derivation here for brevity; the important elements can be found in \citet{Ryden:1970vsj} and \citet{2022ScPA....2....1H}. Since the $H'/H$ factors are related to cosmic time, we computed them in the geometry regime even when $\Om$ was in the growth regime.

We first calculated the linear matter power spectrum with an Einstein-Boltzmann solver (EBS). We call the EBS with the early matter density $\Om^{\rm early}$. Afterwards, the solution of the Einstein-Boltzmann equations was propagated to the late-time Universe with the growth factor. We therefore modelled $G$ in the growth regime with $\Om^{\rm growth}$ and rescaled the linear matter power spectrum by exploiting the dependence on $G(z)^2$,
\begin{equation}
    P_{\rm lin}^{\rm split}(k,z)= P_{\rm lin,EBS}^{\rm early}(k,z)\;\left(\frac{G^{\rm growth}(z)}{G^{\rm early}(z)}\right)^2 \, .
    \label{rescale}
\end{equation}
We calculated the growth factor using Eq.\,(\ref{growth-ode}). In this way, the split linear matter power spectrum depends on $\Om^{\rm early}$ except for the growth factor, where it depends on $\Om^{\rm growth}$. The dependence on the geometry regime will come into play later, for example, through the window functions.

This also implies that the variance of linear matter density perturbations on the scale of $R=8\,h^{-1}\textrm{Mpc}$ had to be rescaled in the same way,
\begin{equation}
    \sigma_8^2(z)=\frac{1}{2\pi^2}\int \mathrm{d} \log k P(k, z) W^2(kR)k^3,
    \label{sigma-8-def}
\end{equation}
\begin{equation}
    \sigma_8^{\rm split}(z)=\sigma_{\rm 8, EBS}^{\rm early}(z)\frac{G^{\rm growth}(z)}{G^{\rm early}(z)}.
    \label{sigma8split}
\end{equation}

Most galaxy survey probes rely on information hidden in small scales, where linear theory breaks down. Since we only rescaled the linear matter power spectrum, we derived the split for the non-linear matter power spectrum. To do this, we used the boost factor. The boost factor quantifies non-linearities in the matter power spectrum that arise from structure formation at small scales and thus concern structure growth,
\begin{equation}
    B^{\rm growth}(k,z)=P_{\rm NL}^{\rm growth}(k,z)/P_{\rm linear}^{\rm growth}(k,z) \, ,
\end{equation}
\begin{equation}
    P_{\rm NL}^{\rm split}(k,z)=P^{\rm split}_{\rm linear}(k,z)\,B^{\rm growth}(k,z) \, .
\end{equation}

With this rescaling and the growth-regime boost, the split matter power spectrum can be calculated using different values for $\Om^{\rm early}$ and $\Om^{\rm growth}$. This is similar to the split matter power spectrum used by \cite{Zhong:2023how}. The differences are that we considered the matter density of the EBS to be in the early regime. Furthermore, in Eq.\,(\ref{growth-ode}), we computed $H$ in the geometry regime, regardless of which $\Om$ we used in the third term of that equation. 

\begin{figure}
	\centering
	\includegraphics[width=\linewidth]{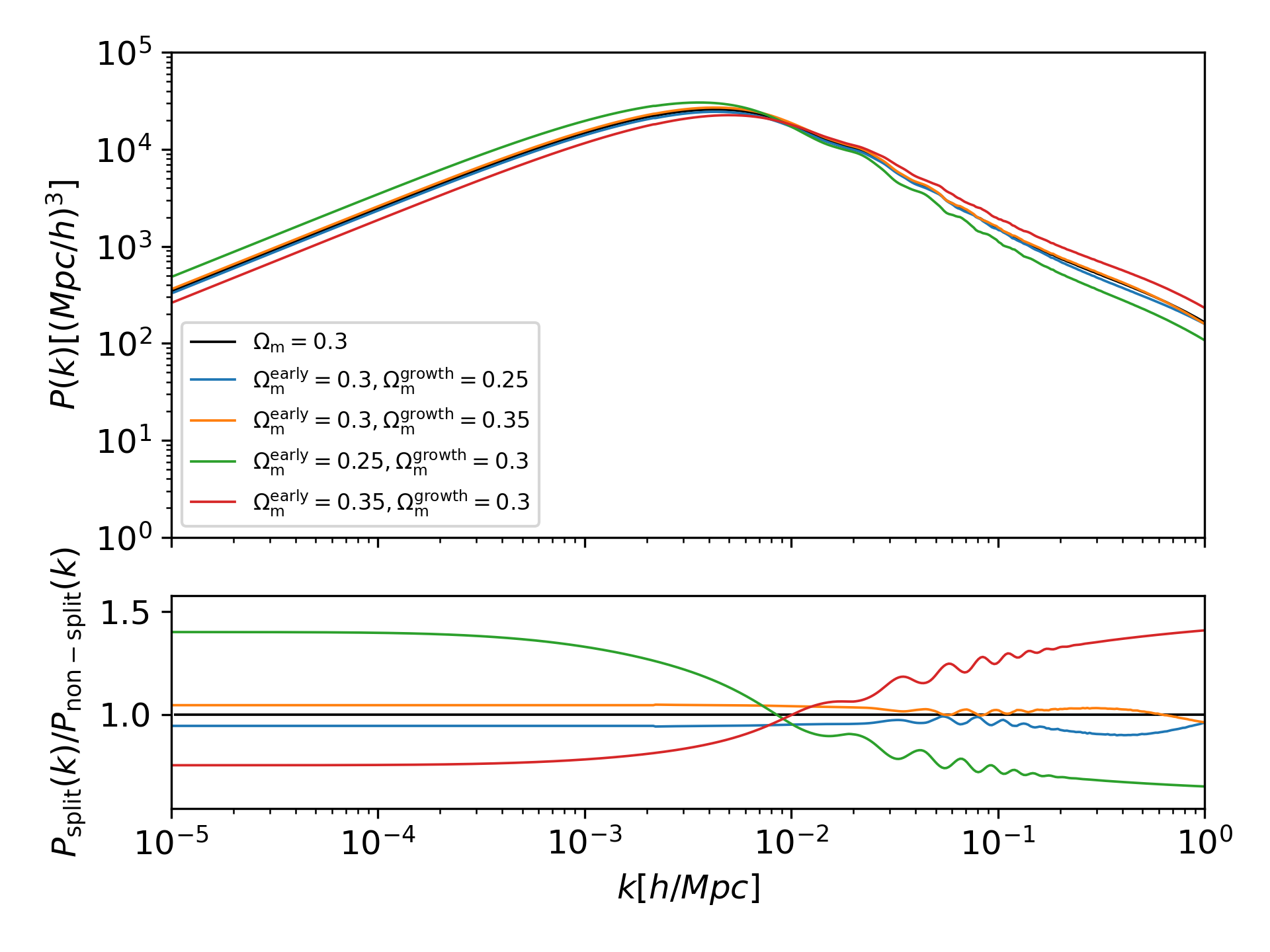}
	\caption{Top: Split non-linear matter power spectrum at redshift $z=0$. Bottom: Ratio of the split non-linear matter power spectrum with respect to the non-split non-linear matter power spectrum at redshift $z=0$ and $\Om=0.3$.}
	\label{fig:Pk_NL_split}
\end{figure}

The split matter power spectrum is shown in Fig.\,\ref{fig:Pk_NL_split}. The early regime affects the matter power spectrum more strongly because the main calculation is based on this parameter. The growth regime rescales the spectrum in a constant way at larger scales, which comes from Eq.\,(\ref{rescale}). At smaller scales, the non-linear boost that is governed by $\Om^{\rm growth}$ is visible.

For the 3x2pt analysis, we considered the angular two-point correlation functions $w^i(\theta)$ for GC, $\gamma_{\rm t}^{ij}(\theta)$ for GGL, and $\xi^\pm_{ij}(\theta)$ for WL,
\begin{gather}
    w_i(\theta)=\sum_\ell C_{ii}^{\delta_{\rm obs} \delta_{\rm obs}}(\ell)\mathcal{G}_0(\ell, \theta_{\rm min}, \theta_{\rm max})\,, \\
    \gamma_{ij}^{\rm t}(\theta)=\sum_\ell C_{ij}^{\delta_{\rm obs} E}(\ell)\mathcal{G}_2(\ell, \theta_{\rm min}, \theta_{\rm max})\,, \\
    \xi_{ij}^{\pm}(\theta)=\sum_\ell \left(C_{ij}^{EE}(\ell)\pm C^{\rm BB}_{ij}(\ell)\right)\mathcal{G}_{4, \pm}(\ell, \theta_{\rm min}, \theta_{\rm max})\,.
\end{gather}

Here, $\theta$ denotes the angular distance, and the analytic functions $\mathcal{G}_n$ can be found in \cite{DES:2020ypx}. The indices $i$ and $j$ refer to the tomographic redshift bins. These angular power spectra have multiple components, as detailed in \cite{PhysRevD.105.023520},

\begin{align}
C^{\delta_{\rm obs} \delta_{\rm obs}}_{ii}(\ell) &= C^{\delta_g \delta_g}_{ii}(\ell) + C^{\delta_\mu \delta_\mu}_{ii}(\ell) + C^{\delta_{\rm{RSD}} \delta_{\rm{RSD}}}_{ii}(\ell),
\label{angular-power-spectra-gc} \\
C^{\delta_{\rm obs E}}_{ij}(\ell) &= C^{\delta_g \kappa}_{ij}(\ell) + C^{\delta_g \rm{I_E}}_{ij}(\ell) + C^{\delta_\mu \kappa}_{ij}(\ell) + C^{\delta_\mu \rm{I_E}}_{ij}(\ell) \nonumber\\ &+ 2C^{\delta_g \delta_\mu}_{ii}(\ell) + 2C^{\delta_g \delta_{\rm RSD}}_{ii}(\ell) + 2C^{\delta_{\rm RSD} \delta_\mu}_{ii}(\ell), 
\label{angular-power-spectra-xc} \\
C^{\rm EE}_{ij}(\ell) &= C^{\kappa \kappa}_{ij}(\ell) + C^{\rm \kappa I_E}_{ij}(\ell) + C^{\kappa \rm I_E}_{ji}(\ell) + C^{\rm I_E I_E }_{ij}(\ell), \label{angular-power-spectra-EE} \\
C^{\mathrm{BB}}_{ij}(\ell) &= C^{\mathrm{I_B I_B}}_{ij}(\ell)\,,
\end{align}

where $\delta_{\rm obs}$ stands for the observed galaxy number counts and includes contributions from the density, $\delta_g$, magnification, $\delta_{\mu}$, and RSD, $\delta_{\rm RSD}$. E and B represent the E- and B-modes of cosmic shear, which include contributions from matter, $\kappa$, and intrinsic alignments (IA), $I_{\rm E}$ and $I_{\rm B}$. As can be seen in Eq. \ref{angular-power-spectra-gc}, for GC only the auto-correlation functions are taken into account.

We used the split matter power spectrum to calculate the angular power spectra $C^{AB}_{ij}(\ell)$, where the superscript labels $A$ and $B$ refer to the probes that were used. These are integrals over window functions for the redshift bins $i$ and $j$ from 0 until the comoving distance to the last scattering surface ($\chi_{\rm LSS}$),

\begin{align}
    C_{ij}^{AB}(\ell)=\int_0^{\chi_{\rm LSS}} &\frac{W^A_i(\chi)W_j^B(\chi)}{\chi^2} \, \nonumber\\
    &\times P_{\rm NL}^{\rm split}\left(k=\frac{\ell+\frac{1}{2}}{\chi},z(\chi)\right)\text{d}\chi\, .
    \label{angular-ps-general}
\end{align}

This assumes the Limber approximation, which was used except for GC, the full formalism of which is detailed in \citep{Fang:2019xat}. The magnification terms in Eqs. \ref{angular-power-spectra-gc} and \ref{angular-power-spectra-xc} come from the gravitational lensing of the galaxies. This should be treated in the same way as the WL window function and the prefactor in the window function is thus geometric (see Appendix \ref{appendix-wl-window}). In the RSD contribution to GC, the growth factor $f(z)$ appears in the window function since RSD is governed by structure growth. We used the growth factor from the differential equation, Eq.\,(\ref{growth-ode}), using $\Om^{\rm growth}$.

In the case of GC, the window function depends on geometry through the comoving distance $\chi$ \citep[see, e.g.,][]{DES:2021zxv},
\begin{equation}
    W_i^{\rm \delta g}(\chi)=b_i(z(\chi))  \frac{n_i^{g}(z(\chi))\text{d}z/\text{d}\chi}{\bar{n}^g_i}\,,
    \label{gc-window}
\end{equation}
where $b_i$ stands for the linear galaxy bias, $n_i^g$ represents the galaxy redshift distribution, and $\bar{n}^g_i$ corresponds to the mean galaxy redshift distribution.

\begin{figure}
	\centering
	\includegraphics[width=\linewidth]{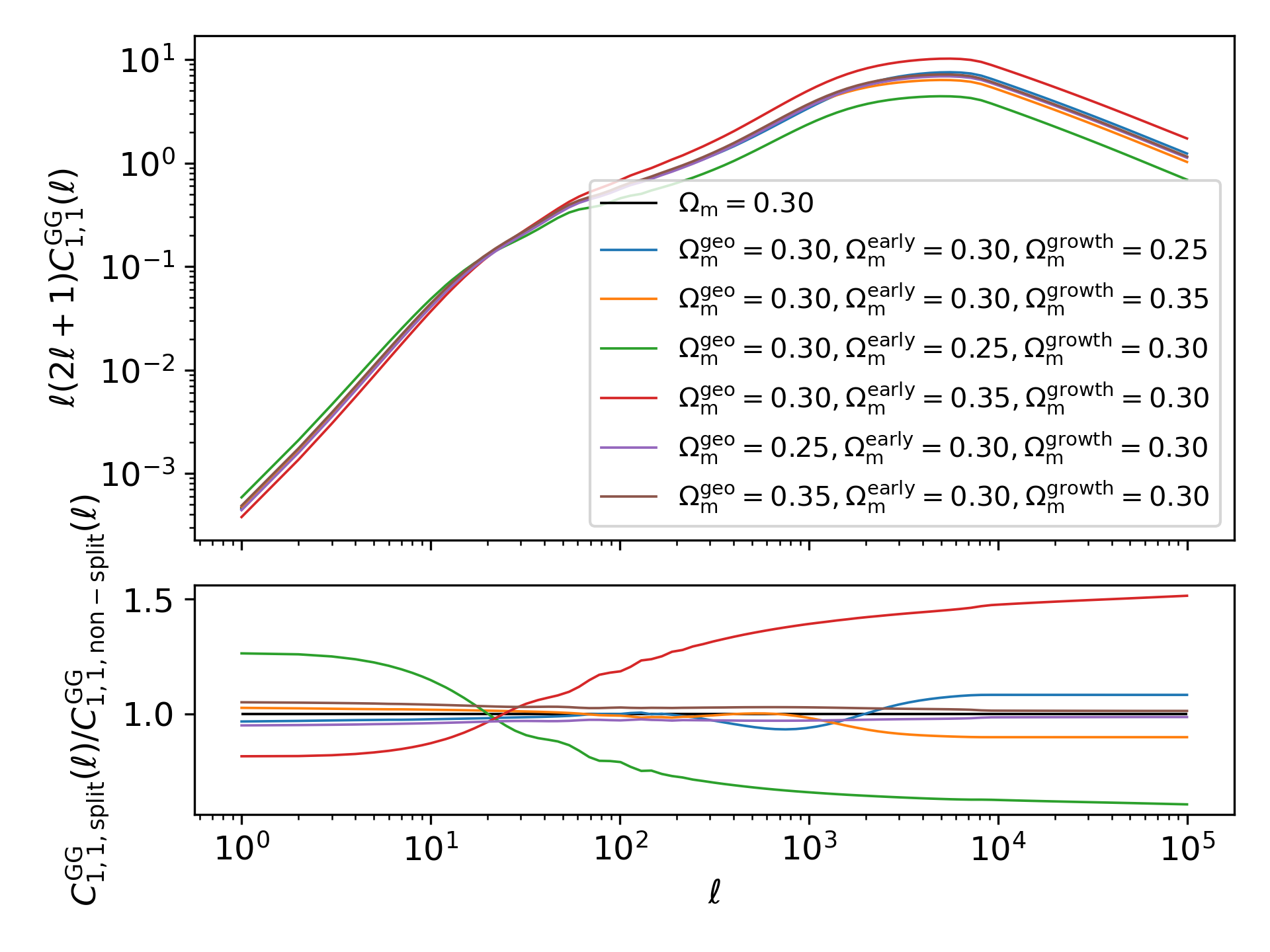}
	\caption{Top: Split angular power spectrum of GC for the auto-correlation in the first bin (1,1) with different values for $\Om$ in the three regimes, showing the dependence on geometry, growth, and the early Universe. Bottom: Ratio of the split angular power spectrum of GC with respect to the non-split angular power spectrum at $\Om=0.3$.
    }
	\label{fig:C_ell_GC}
\end{figure}

The GC angular power spectrum thus depends on the early and growth regimes through the split matter power spectrum and on geometry through the window function. The dependence is shown in Fig.\,\ref{fig:C_ell_GC}. We show the angular power spectrum for different values of $\Om$ in the three regimes. We focus on the auto-correlation in the first bin according to DES Y3 specifications with bin edges $0.15<z<0.35$. This bin combination has the highest signal-to-noise ratio (S/N). In the bottom plot, the ratios with respect to the standard non-split $C^{\rm GG}_{\rm non-split}(\ell)$ are shown. The early regime (in green and red) has the strongest effect because the matter power spectrum is based on the early regime, and this constitutes the basis for the angular power spectrum.

The observed shear contains the WL measurement ($\kappa$), but also the IA of galaxies ($I_B, I_E$), which contaminates the measurement \citep[see, e.g.,][]{Bridle:2007ft}, as shown in Eq.\,(\ref{angular-power-spectra-EE}).  For the gravitational shear, in addition to the redshift distribution, the window function for $\kappa$ contains $\Om$ in the prefactor \citep[see, e.g.,][]{Kilbinger:2017lvu}. In the derivation, this comes from the background matter density (see Appendix \ref{appendix-wl-window}), so we classified this as a geometric quantity, 
\begin{equation}
    W_i^{\kappa}(\chi)=\frac{3H_0^2\Om^{\text{geo}}}{2c^2}\frac{\chi}{a(\chi)}\int_\chi^\infty \text{d}\chi' \frac{n_i^\kappa(z(\chi'))\text{d}z/\text{d}\chi'}{\bar{n}^\kappa_i}\frac{\chi'-\chi}{\chi}\,.
    \label{wl-window}
\end{equation}

Here, $n_i^{\kappa}$ is the redshift distribution for the sources.
In addition to the WL angular spectra, $C^{\rm \kappa \kappa}_{ij}(\ell)$, we computed the IA auto-correlation $C^{\rm I I}_{ij}(\ell)$, and the cross-correlation term, $C^{\rm \kappa I}_{ij}(\ell)$ \citep[see, e.g.,][]{Hirata:2004gc, Campos:2022gtg}. The WL angular spectra are given by Eq.\,(\ref{angular-ps-general}) using the WL window function from Eq.\,(\ref{wl-window}). The II spectra contain the IA power spectrum $P^{\rm II}(k,z)$, 
\begin{equation}
    C^{\rm I I}_{ij}(\ell)=\int d\chi \frac{n_i^\kappa(\chi)n^\kappa_j(\chi)}{\chi^2}P^{\rm II}(k, z(\chi))\,.
\end{equation}

The cross-correlation between the $\kappa$ and the IA terms contains the combined power spectrum $P^{\rm \kappa I}(k,z)$,
\begin{equation}
    C^{\rm \kappa I}_{ij}(\ell)=\int d\chi \frac{W_i^\kappa(\chi)n^\kappa_j(\chi)}{\chi^2}P^{\rm \kappa I}(k, z(\chi))\,.
\end{equation}

The IA power spectra can be modelled with an amplitude that depends on redshift. We used the non-linear alignment model \citep[NLA,][]{Bridle:2007ft}. Our split non-linear matter power spectrum became the basis for the IA power spectra,
\begin{equation}
    P^{\rm I I}(k, z(\chi))=C_{\rm NLA}^2(z)P^{\rm split}_{\rm NL}(k, z(\chi))\,,
\end{equation}
\begin{equation}
    P^{\rm \kappa I}(k, z(\chi))=C_{\rm NLA}(z)P^{\rm split}_{\rm NL}(k, z(\chi))\,.
\end{equation}

Since the amplitude of IA depends on structure growth at small scales, we calculated this amplitude in the growth regime,
\begin{equation}
    C_{\rm NLA}(z)=-\frac{A_1\bar{C} \rho_{\rm crit}\Om^{\rm growth}}{aG(z)}\left(\frac{1+z}{1+z_0}\right)^{\eta_1}.
\end{equation}

Here, $\rho_{\rm crit}$ is the critical density, $z_0$ is the mean redshift of the source galaxy sample, and $\bar{C}$ is a constant ($\bar{C}=5\times10^{-14}h^{-2}M_\odot^{-1}\rm{Mpc}^3$). The NLA model has two degrees of freedom: $A_1$ and $\eta_1$.

\begin{figure}
	\centering
	\includegraphics[width=\linewidth]{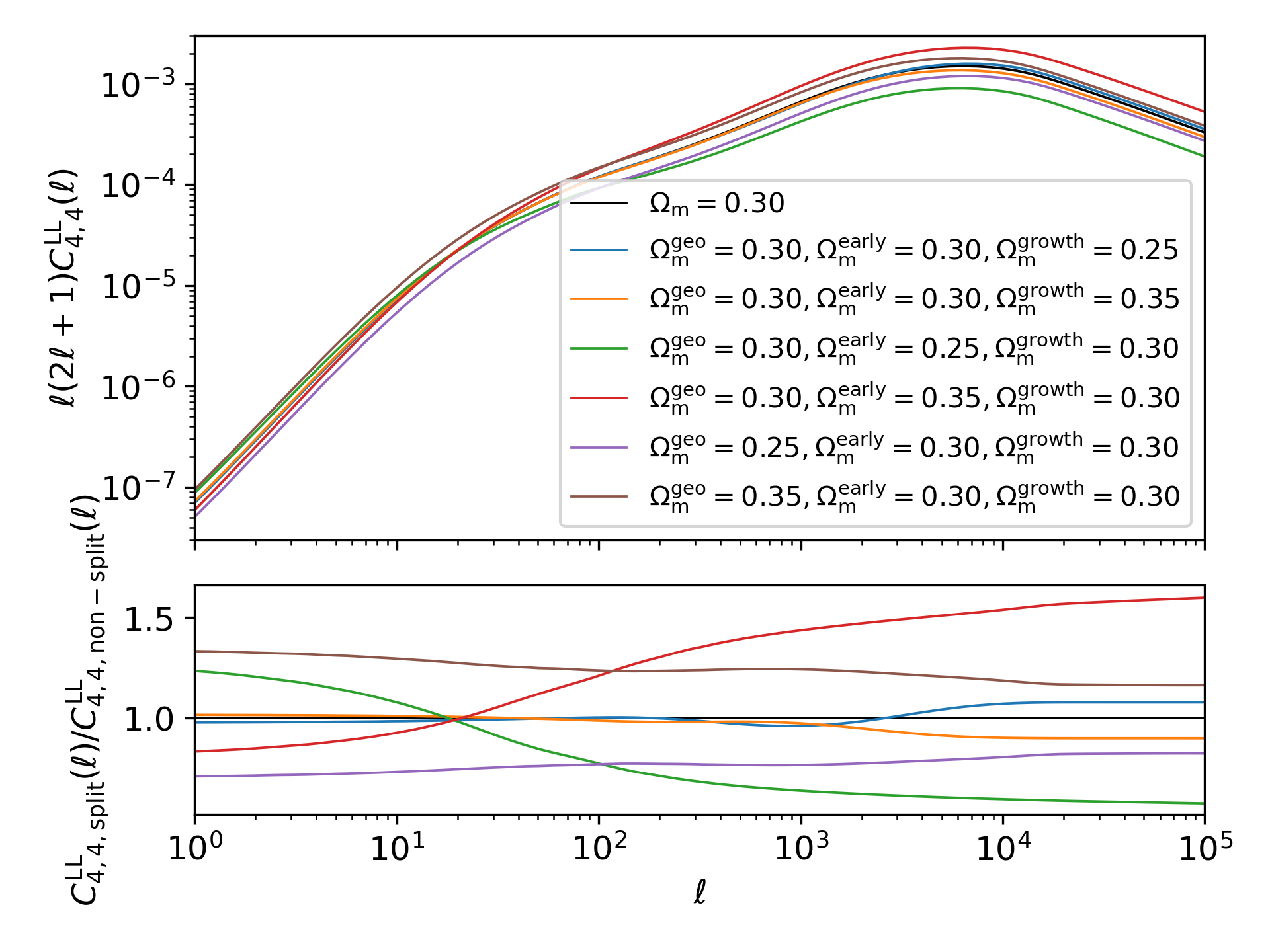}
    \caption{Top: Split angular power spectrum of WL for the auto-correlation in the last DES Y3 bin (4,4) with different values for $\Om$ in the three regimes, showing the dependence on geometry, growth, and the early Universe. Bottom: Ratio of the split angular power spectrum of WL with respect to the non-split angular power spectrum at $\Om=0.3$.}
	\label{fig:C_ell_WL}
\end{figure}

Through the matter power spectrum, the window function, and IA, the WL angular power spectrum also depends on all three regimes, as shown in Fig.\,\ref{fig:C_ell_WL}. We also show the bin combination with the highest S/N. In the WL case, this is the auto-correlation of the bin with the highest redshift. Since we used DES Y3 specifications for these figures, this is the fourth bin, which corresponds to $0.87<z<2.0$.

The angular power spectrum has fewer wiggles than in Fig.\,\ref{fig:C_ell_GC} because the WL window function contains an integral along the line of sight. Furthermore, we evaluated the matter power spectrum at a higher redshift, which also reduces the wiggles.
The effect of the geometric regime (in purple and brown) is stronger than in Fig.\,\ref{fig:C_ell_GC}. This comes from the WL window function, see Eq.\,(\ref{wl-window}), which contains $\Om^{\rm geo}$.

The 3x2pt data also include the cross-correlation between GC and WL using the angular power spectrum from Eq.\,(\ref{angular-ps-general}). This is shown in Appendix \ref{appendix-ggl}.

\subsubsection{Cosmic microwave background}
\label{sect:cmb}
We computed the CMB angular power spectrum with an EBS in the early-universe regime, as we did for the 3x2pt observables. CMB observables mostly reflect early-universe physics, but they also contain information about geometry, which comes from the distance to the last scattering surface of photons. 

The CMB measures the angular scale of acoustic peaks $\alpha$. This is the ratio of the sound horizon at the time of decoupling and the angular diameter distance, also at decoupling,
\begin{equation}
    \alpha = \frac{r_{\rm s}(\eta_{\rm dec})}{d_{\rm A}(\eta_{\rm dec})}=\frac{\int_0^{\eta_{\rm dec}}d\eta\;c_{\rm s}(\eta)}{d_{\rm A}(\eta_{\rm dec})}\,.
    \label{theta-def}
\end{equation}

Here, the numerator is an integral of the sound speed $c_{\rm s}$ from conformal time $\eta=0$ until decoupling. This means that it should be calculated in the early regime. On the other hand, the denominator contains the angular diameter distance, which is in the geometry regime.

Each multipole moment $\ell$ corresponds to a configuration in which a maximum and a minimum are separated by the angle $\alpha = \pi/\ell$. Our original multipole space was calculated in the early regime. To account for the angular diameter distance in Eq.\,(\ref{theta-def}), we therefore rescaled the multipole axis of our computed $C(\ell)$,
\begin{equation}
    C^{\rm TT/TE/EE}(\ell_{\rm rescaled})=C^{\rm TT/TE/EE}\left(\ell_{\rm EBS}\frac{d_{\rm A}^{\rm geo}}{d_{\rm A}^{\rm early}}\right)\,,
\end{equation}
where T stands for temperature and E for polarisation.

The split CMB probe is therefore sensitive to the early Universe and also to geometry through this rescaling. We note that, in principle, the CMB probe should also be sensitive to the growth of perturbations via the effect of gravitational lensing, for example. However, we defer the inclusion of these effects to future work and focus here on the primordial CMB and our distance to it.
In Sect.\,\ref{planck} we describe how we selected the CMB data to remove those that were most sensitive to the growth regime.

\subsubsection{Baryon acoustic oscillations}
The BAO measurements are based on galaxy surveys and measure statistics of separation between pairs of galaxies. Two quantities can be observed with this probe. For distances between galaxies that are perpendicular to the line of sight, the observed quantity is $\Delta \beta=r_{\rm d}/ d_{\rm M}$. This angle corresponds to the sound horizon at the drag epoch shortly after decoupling, $r_{\rm d}$, divided by the comoving distance $d_{\rm M}$.

When the pair of galaxies is separated along the line of sight, this results in a difference in redshift $\Delta z=r_{\rm d}/d_{\rm H}$. When the S/N is too low, the ratio of the angle-averaged distance $d_{\rm V}$ and the sound horizon can be measured,
\begin{equation}
    d_{\rm V}=(z d^2_{\rm M}(z)d_{\rm H}(z))^{1/3}\,.
\end{equation}

The same reasoning applies to other tracers, such as quasars.
We used $\Om^{\rm geo}$ for the distances, that is, the Hubble distance $d_{\rm H}$, the comoving distance $d_{\rm M}$, and the angular diameter distance $d_{\rm V}$. These distances are not sensitive to structure formation or early-universe physics. 

The sound horizon at the drag redshift is an integral over the sound speed $c_{\rm s}$, as in Eq.\,(\ref{theta-def}). The integral concerns times from the Big Bang to the baryon drag epoch, making it an early-universe quantity,
\begin{equation}
    r_{\rm d} = r_{\rm s}(z_{\rm d}) = \int_{z_{\rm d}}^{\infty}\frac{c_{\rm s}(z)}{H(z)}\mathrm{d}z\,.
\end{equation}

In practice, we calculated it using the approximation by \cite{Brieden:2022heh} and $\Om^{\rm early}$,
\begin{align}
    r_{\rm d}\approx & \, 147.05\,\mathrm{Mpc}\nonumber\\
    &\times\left(\frac{\Omega_{\rm m}^{\rm early}h^2}{0.1432}\right)^{-0.23}\left(\frac{N_{\rm{eff}}}{3.04}\right)^{-0.1}\left(\frac{\Omega_{\rm b} h^2}{0.0224}\right)^{-0.13}\,.
    \label{r_d_equation_desi}
\end{align}

This approximation depends on the physical baryon density $\omega_{\rm b}=\Omega_{\rm b} h^2$ and the effective number of relativistic species $N_{\rm eff}$. Furthermore, it assumes $\Lambda$CDM and standard early-time physics to estimate $r_{\rm d}$. Since we used $\Om^{\rm early}$ in the early regime to take this into account, this is consistent with our split approach. We did not use $r_{\rm d}$ from the EBS to be able to run this separately without calling an EBS.

\subsubsection{Type Ia supernovae}
With cosmological SNe Ia data, we can directly measure the distance from the observer to the SNe Ia using the luminosity distance $d_{\rm L}$. SNe Ia are standardisable candles, meaning that their absolute magnitude $M_0$ can be standardised and used to infer the luminosity distance when the apparent magnitude $m$ is measured. 

The luminosity distance depends on the cosmological parameters $H_0$ and $\Omega_{\rm m}$. Since this probe is only sensitive to a distance, we classified it into the geometry regime and related the apparent magnitude to the luminosity distance using $\Omega_{\rm m}^{\rm geo}$,
\begin{equation}
    m(z, \Omega_{\rm m}^{\rm geo},H_0)=M_0+5 \log_{10}\left[\frac{d_L(z,\Omega_{\rm m}^{\rm geo},H_0)}{\rm{Mpc}}\right]+25\,.
\end{equation}

\subsubsection{Redshift-space distortions}
The RSD are distortions that are introduced when mapping from real space to redshift space. This happens because the Universe is not homogeneous, and objects such as galaxies have peculiar velocities in addition to their velocity due to the expansion of the Universe (i.e. the Hubble flow; see e.g. \citet{Kaiser:1987qv}). The recession velocity $cz$ of an object depends on the Hubble parameter, on its peculiar velocity, and that of the observer,
\begin{equation}
    cz = H_0 r+[v(r)-v(0)]\,.
\end{equation}

Here, $r$ is the distance from the observer to the object. This makes two-point statistics in clustering analyses sensitive to the product of the growth rate parameter $f(z)$ and $\sigma_8$ \citep[see e.g. ][]{Percival:2008sh, Taruya:2010mx},
\begin{equation}
    f\cdot \sigma_8 (a) := \frac{\rm{d}\ln D(a)}{\rm{d}\ln a} \sigma_8 (a) = \frac{\rm{d}\ln ((1+z)^{-1}G(a)) }{\rm{d}\ln a}\sigma_8 (a)\,.
    \label{f-sigma-8-def}
\end{equation}

We used $f(z)$ calculated from the differential equation in Eq.\,(\ref{growth-ode}), using $\Om^{\rm growth}$ and $\sigma_8^{\rm split}$, as we defined it in Eq.\,(\ref{sigma8split}). 

\section{Data}
\label{sec:data}
\subsection{Dark Energy Survey Y3 -- 3x2pt}
Using the photometric measurements from the stage III galaxy survey DES  \citep{PhysRevD.105.023520, DES:2026fyc}, we performed 3x2pt analyses which consist of three different two-point correlation functions of galaxies. We used the redMaGiC lens sample here, although we show in Appendix\,\ref{appendix-maglim} that the joint analysis using the MagLim sample provides very similar results.

\begin{figure}
	\centering
	\includegraphics[width=\linewidth]{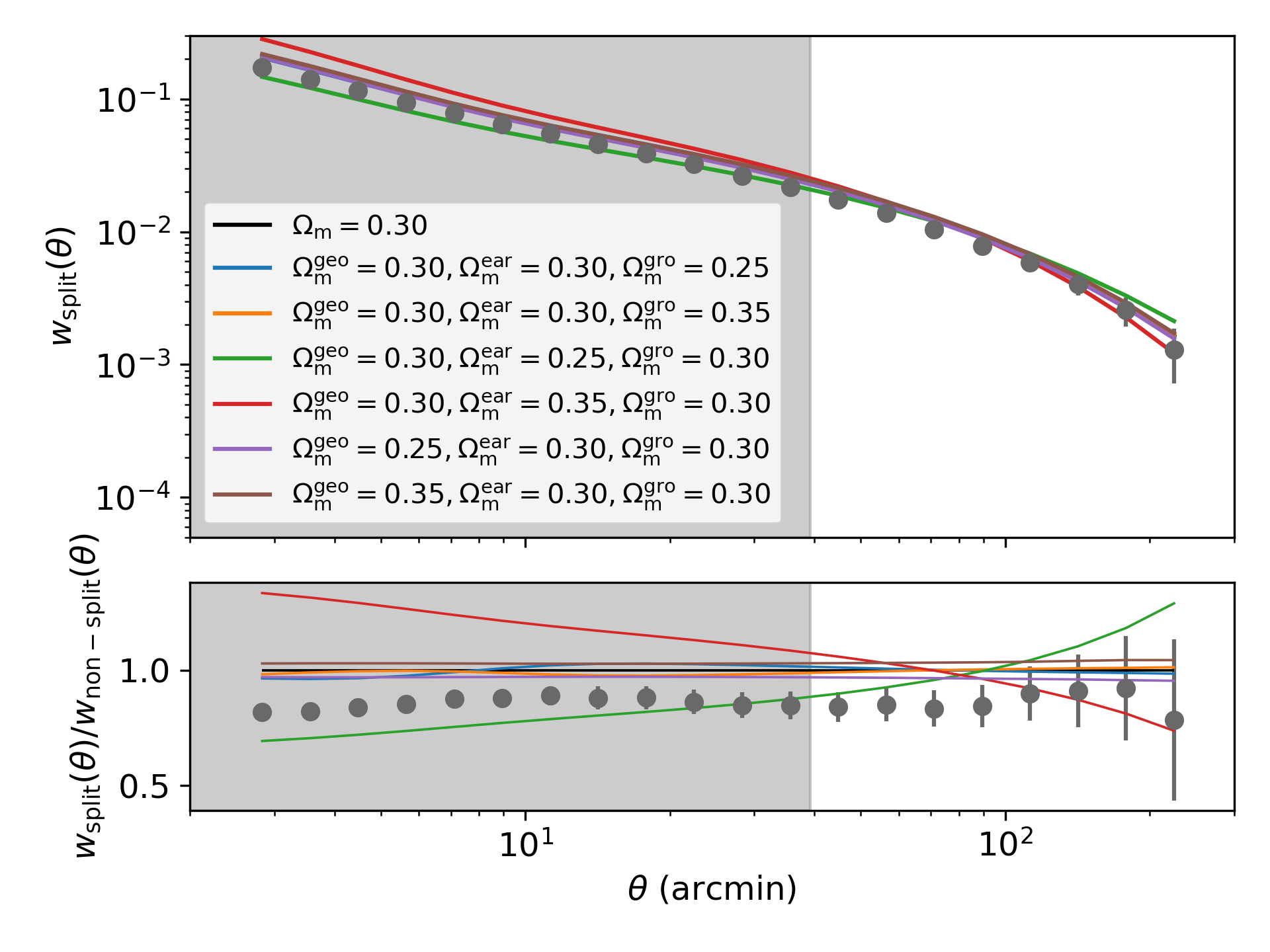}
	\caption{Top: Split two-point auto-correlation function for GC
    with DES Y3 data in the tomographic bin (1,1),
    including theoretical predictions with different $\Om$ values per regime. The DES Y3 data points are shown in grey, and the grey band indicates the scale cuts. Bottom: Ratio of the split correlation function with respect to the non-split correlation function with $\Omega_{\rm m}=0.3$.}
	\label{fig:w-theta}
\end{figure}

One probe is GC using the lens galaxies with the angular correlation function $w_{ij}(\theta)$. In the DES analysis, only the auto-correlations $w_{ii}(\theta)$ were used. Out of these, the first bin ($0.15<z<0.35$) has the highest S/N. We show the DES Y3 data and how they compare to theoretical predictions with different values for $\Om$ per regime in Fig.\,\ref{fig:w-theta}. Here and in all further plots that show the predictions of the observables, we used the best-fit values from \textit{Planck} 2018 \cite{Planck:2018vyg}, except for the indicated values for $\Om$. The most influential regime here is the early regime because this governs the main part of the matter power spectrum. The effect of this is a tilt because $A_{\rm s}$ remains constant in all the predictions. 

\begin{figure}
	\centering
	\includegraphics[width=\linewidth]{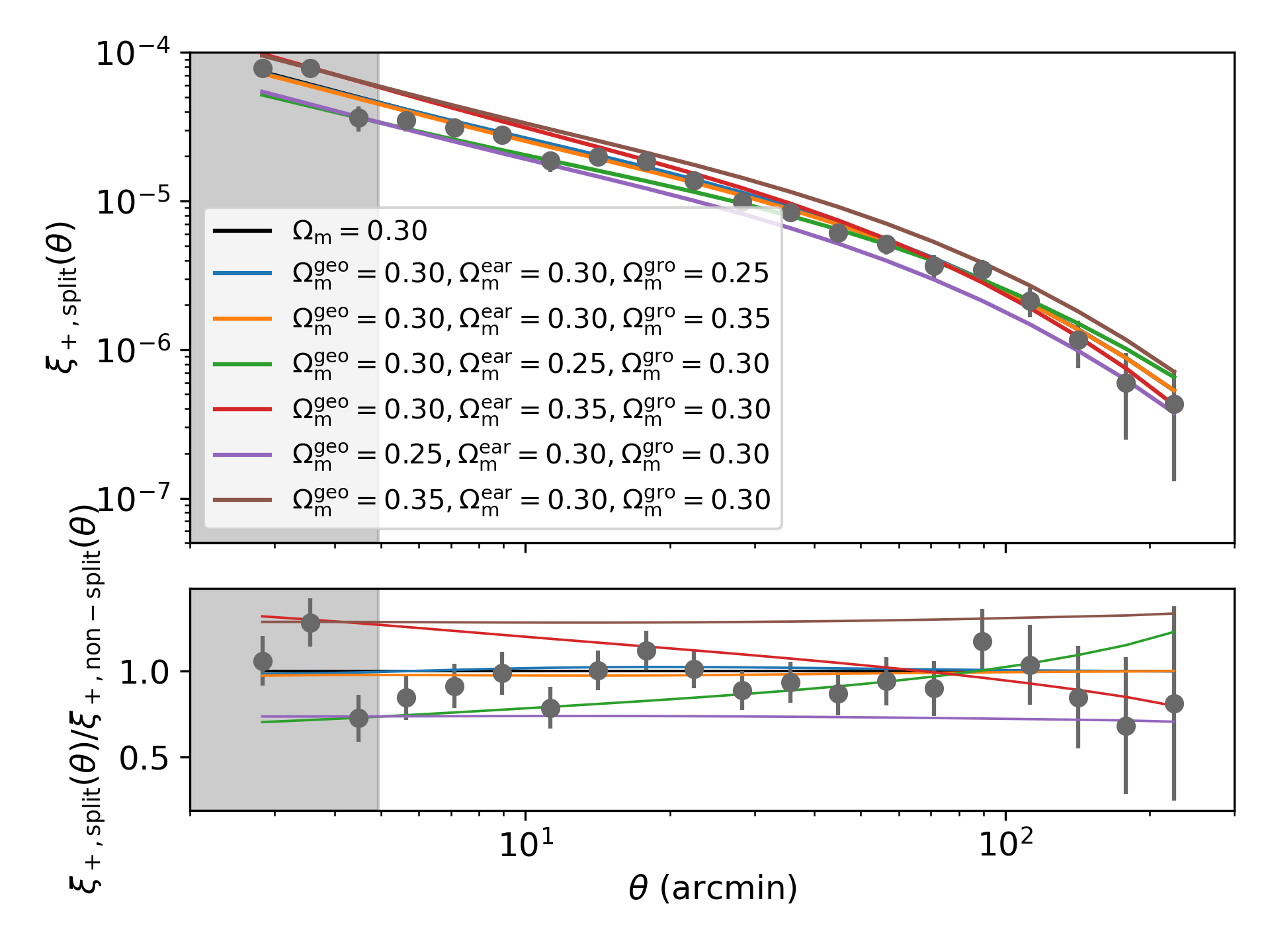}
	\caption{Top: Split two-point correlation function for WL
    with DES Y3 data in the tomographic bin (4,4), including theoretical predictions with different $\Om$ values per regime. The DES Y3 data points are shown in grey, and the grey band indicates the scale cuts. Bottom: Ratio of the split correlation function with respect to the non-split correlation function with $\Omega_{\rm m}=0.3$.}
	\label{fig:xi-plus}
\end{figure}

The two-point correlation functions $\xi_{ij}^\pm(\theta)$ describe the WL correlation between galaxies in bins $i$ and $j$. $\xi_{ij}^+$ describes the sum and $\xi_{ij}^-$ describes the difference of the products of the tangential and cross-components. To illustrate the theoretical predictions using the three different regimes, we show the DES Y3 measurements of the $\xi^+(\theta)$ function in bin (4,4) ($0.87<z<2.0$), again without scale cuts, in Fig.\,\ref{fig:xi-plus}. This bin combination has the highest S/N of the bins used in the analysis. The effect of the growth regime is weak because this would become important at smaller scales. The early regime represents a tilt in the prediction, whereas the geometric regime shifts the prediction due to the window function.

The cross-correlation between GC and WL is described by the GGL two-point correlation function $\gamma^{\rm t}_{ij}(\theta)$. The theoretical predictions compared to the data are shown in Appendix \ref{appendix-ggl}.

When performing the parameter inference, we employed the same scale cuts as in the DES Y3 analysis (see \cite{PhysRevD.105.023520}). We varied the same parameters as in the non-split analysis (see Appendix \ref{nuisance-des}).

\subsection{\textit{Planck} 2018 cosmic microwave background}
\label{planck}
For CMB anisotropies, we used temperature, polarisation, and their cross-correlation (TT, TE, and EE) power spectra from the \textit{Planck} 2018 measurements \citep{Planck:2018vyg, Planck:2019nip}. In addition to geometry and early-universe information, the CMB also contains information about structure growth. In our modelling, we focused on separating the geometry and early-universe information of the CMB. To factor out the influence of growth, we only used a subset of the \textit{Planck} data. 

We did not include the lensing power spectra because they are affected by structure growth. Furthermore, we followed the approach of \cite{Zhong:2023how} and applied scale cuts to the high-$\ell$ TT, TE, and EE power spectra. This removed scales after the first acoustic peak and kept scales at $35<\ell<396$. Since lensing affects the CMB temperature spectrum more strongly at smaller scales where there is more structure growth \citep{Hanson:2009kr}, this removes lensing information in the TT spectrum.

For low-$\ell$ likelihoods ($\ell<30$), we only used EE polarisation spectra. In this configuration, contributions from the integrated Sachs-Wolfe effect to the temperature power spectrum are removed. This is a growth effect because it comes from CMB photons passing through gravitational structures that affect their energy, and is important at small $\ell$ \citep[see, e.g.,][]{Weinberg:2008zzc}. Regarding the nuisance parameters, we varied the same parameters as in the non-split analysis (see Appendix \ref{nuisance-planck}).

\subsection{Dark Energy Spectroscopic Instrument Data Release 2, baryon acoustic oscillations}

\begin{figure}[h]
	\centering
	\includegraphics[width=\linewidth]{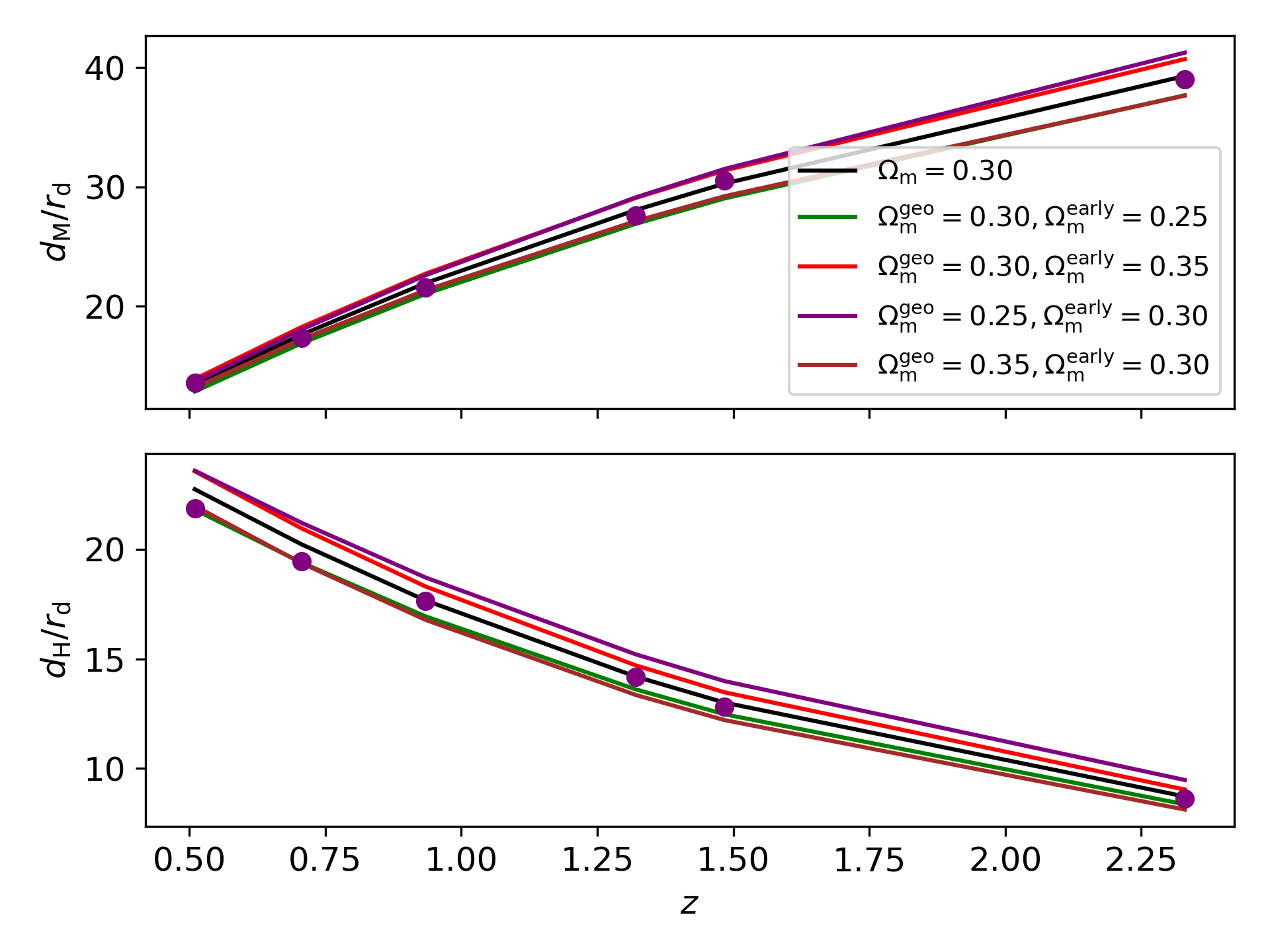}
	\caption{Top: Comoving distance divided by the sound horizon at the drag redshift, with different $\Om$ values per regime. The DESI DR2 data points are shown in purple. Bottom: Hubble distance divided by the sound horizon at the drag redshift.}
	\label{fig:bao-dv}
\end{figure}

For the BAO data, we used the recent DESI DR2 data release \citep{DESI:2025zgx}. It contains $\approx$ 14 million galaxies and quasars in total. These lie in the redshift range of $0.1<z<4.2$. The seven tracer types are the bright galaxy survey, luminous red galaxies, emission line galaxies, quasars, and Lyman-$\alpha$ data.

We used the angular distance and the Hubble distance, each divided by the sound horizon at the drag redshift ($r_{\rm d}$), except for the first bin, where only the angle-average distance divided by $r_{\rm d}$ could be measured. In Fig.\,\ref{fig:bao-dv} we show the theoretical predictions of $d_{\rm M}/r_{\rm d}$ and $d_{\rm H}/r_{\rm d}$ for different values of $\Om$ in the three regimes together with the DESI DR2 data points. The error bars are too small to be visible in this figure. The geometric regime has the strongest effect since the distances relate to this regime. The early regime also affects the prediction, which comes from the sound horizon.

\subsection{Pantheon+ type Ia supernovae}
\label{Pantheon}
We used the Pantheon+ dataset \citep{Scolnic:2021amr}, which contains 1701 SNe Ia. To compute the likelihood, we used the version implemented in \texttt{CosmoSIS} \citep{Zuntz:2014csq}. This uses the Tripp 1998 corrected magnitude \citep{Tripp:1997wt}.

The Pantheon+ dataset includes the SH0ES data for Cepheid calibration. This essentially fixes the absolute magnitude $M_0$ and thus anchors the Hubble diagram at low redshifts. When this anchoring is used, the dataset constrains the Hubble parameter to $73.5\pm 1.1 \, \kmsMpc$ \citep{Brout:2022vxf}. Under the assumption of a $\Lambda$CDM model, this differs from CMB data, for example from \textit{Planck} 2018, producing the Hubble tension. For this reason, we used Pantheon+ without the SH0ES calibration, in which case the SNe Ia sample does not constrain $H_0$. In this configuration, a redshift cut ($z>0.01$) is applied due to systematic uncertainties in the dataset \citep{Brout:2022vxf}. We varied the absolute magnitude $M_0$ as a nuisance parameter in our analysis.

\subsection{Collection of redshift-space distortions measurements}
When we considered RSD measurements, we typically used one tracer sample containing many objects to infer one value of $f\sigma_8$ at an effective redshift $z_{\rm eff}$. \cite{Blanchard:2022xkk} compiled 23 RSD measurements from different spectroscopic galaxy surveys. These measurements were regarded as independent since they differ in tracer object, redshift bin, or sky area. The surveys they used were 2MTF \citep{Howlett:2017asq}, 6dFGS \citep{Jones:2009yz}, GAMA \citep{10.1093/mnras/stv1436}, WiggleZ \citep{Blake:2011rj}, BOSS \citep{Lange:2021zre}, eBOSS \citep{eBOSS:2018abz}, VIPERS \citep{delaTorre:2016rxm}, and Fastsound \citep{Okumura:2015lvp}.

\begin{figure}[h]
	\centering
	\includegraphics[width=\linewidth]{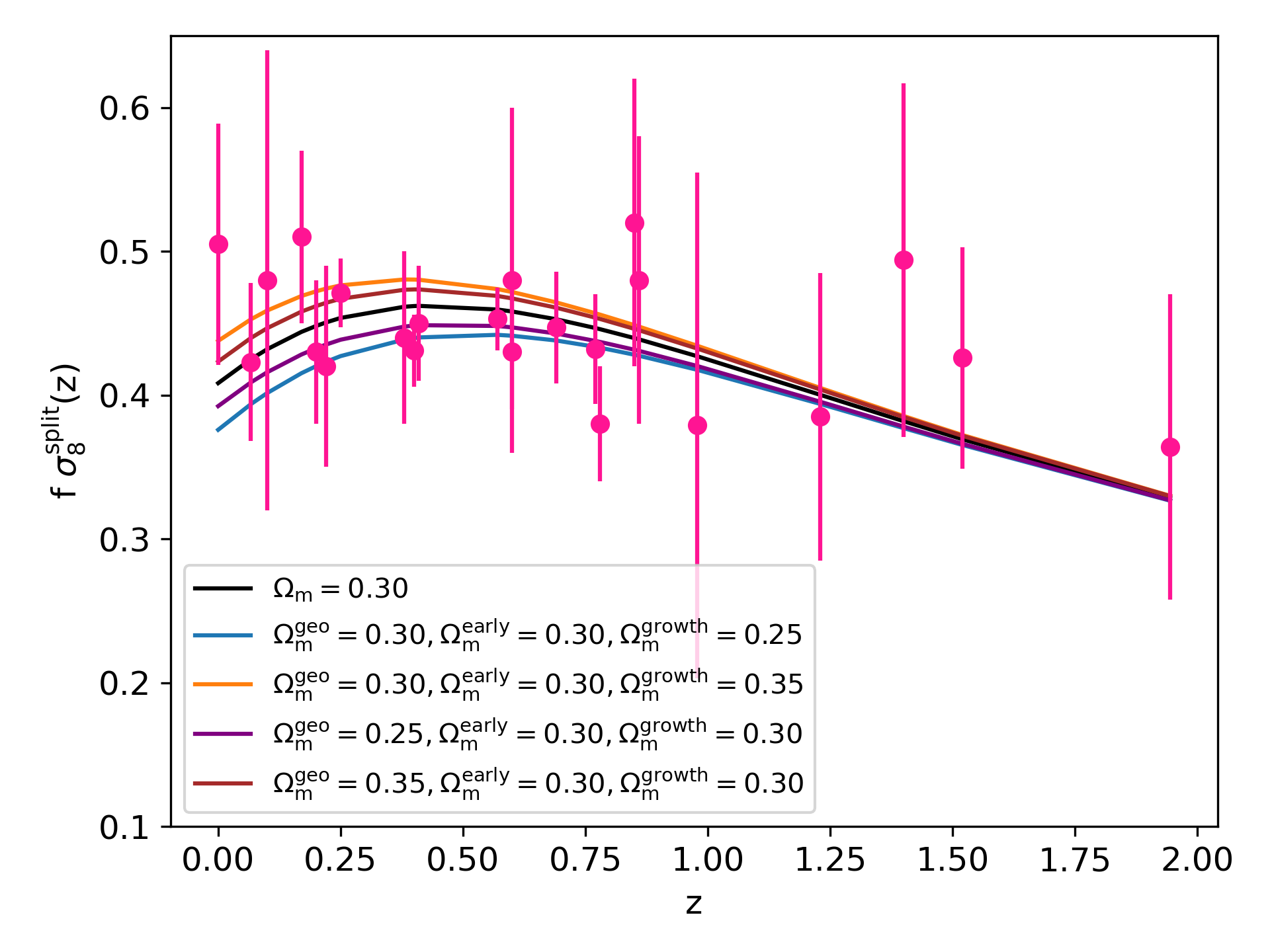}
	\caption{All $f \sigma_8$ data points taken from the RSD measurements compiled in \cite{Blanchard:2022xkk} are shown in pink. The theory predictions are shown for different values of $\Om$ per regime.}
	\label{fig:rsd-data}
\end{figure}

We show the data points together with the $f\sigma_8^{\rm split}(z)$ prediction, using $\sigma_8^{\rm split}$ from Eq.\,(\ref{sigma8split}), in Fig.\,\ref{fig:rsd-data}.\footnote{We do not plot different values of $\Om^{\rm early}$ here, since we focus on the regimes that are physically relevant for RSD.} The growth regime changes the predicted $f\cdot \sigma_8^{\rm split}(z)$ values at lower redshifts. Moreover, the geometric regime affects this to a lesser extent, originating from the factors $H'/H$ in the differential equation (Eq.\,\ref{growth-ode}).

Even though the galaxy samples of these surveys and DESI overlap, we considered the covariance to be negligible here, especially since the observed quantities are very different in character. We know that even the correlation between the pre- and post-reconstruction power spectra is small in practice because of non-linearity, shot noise, and finite scale ranges \citep{Maus:2026wsb}.

\section{Likelihood sampling}\label{sec:likelihood}
To sample the parameter space, we used a modified version of \texttt{CosmoSIS}.
The linear matter power spectrum was computed with \texttt{CAMB} \citep{Lewis:1999bs, Howlett:2012mh} and rescaled using Eq.\,(\ref{rescale}). For the non-linear matter power spectrum, we used \texttt{EuclidEmulator2} \citep{Euclid:2020rfv}. For the combination of all probes, the parameter range of \texttt{EuclidEmulator2} proved to be sufficient because the probes constrain all cosmological parameters within the prior. The effects of this prior are visible in some posterior distributions, such as for $\Om^{\rm growth}$ in the scale-cut \textit{Planck} likelihood. This is not a problem in this analysis because we are interested in the joint constraint and it is not necessary that every observable constrains every parameter.

The pipeline is based on the DES Y3 analysis pipeline in \texttt{CosmoSIS} used to obtain the cosmological constraints in \cite{PhysRevD.105.023520}. This implies a consistency condition on the relation between the physical baryon density $\omega_{\rm b}$, the helium abundance $Y_{\rm He}$, and the number of relativistic degrees of freedom $N_{\rm eff}$ \citep{Pisanti:2007hk}. It further uses the algorithm \texttt{Fast-PT} \citep{McEwen:2016fjn, Fang:2016wcf} and an analytical marginalisation to mitigate non-local effects for GGL \citep{MacCrann:2019ntb}. We also varied the summed neutrino mass $\sum m_\nu$ and show the posteriors that were marginalised over this parameter.

We used the nested sampler \texttt{Nautilus} \citep{Lange:2023ydq}, which employs deep neural networks to efficiently sample the parameter space. This is faster than traditional algorithms such as a Markov chain Monte Carlo method. To combine posteriors from different datasets, we used \texttt{CombineHarvesterFlow} \citep{Taylor2024CombineHarvesterFlow} in the testing phase. This tool uses normalising flows to combine posteriors for non-covariant samples. However, to be as accurate as possible, we used nested sampling on all probes combined for the final results shown below. We ran \texttt{Nautilus} with four neural networks and 100 likelihood evaluations per step. We used 1000 live points when sampling a single probe and 2000 live points when sampling all probes combined.

\section{Results}
\label{sec:results}

\begin{figure}[h]
	\centering
    \begin{subfigure}{0.35\textwidth}
        \includegraphics[width=\linewidth]{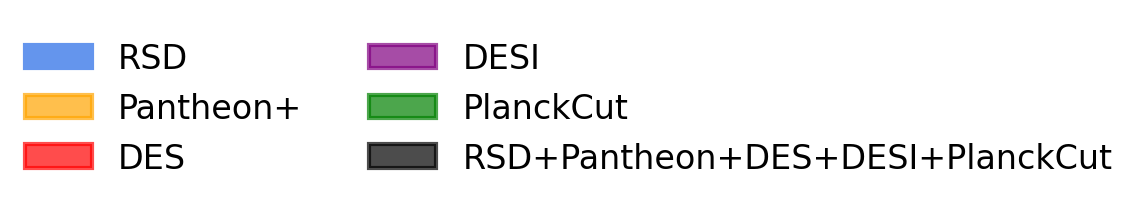}
    \end{subfigure}
	\begin{subfigure}{0.24\textwidth}
    	\includegraphics[width=\linewidth]{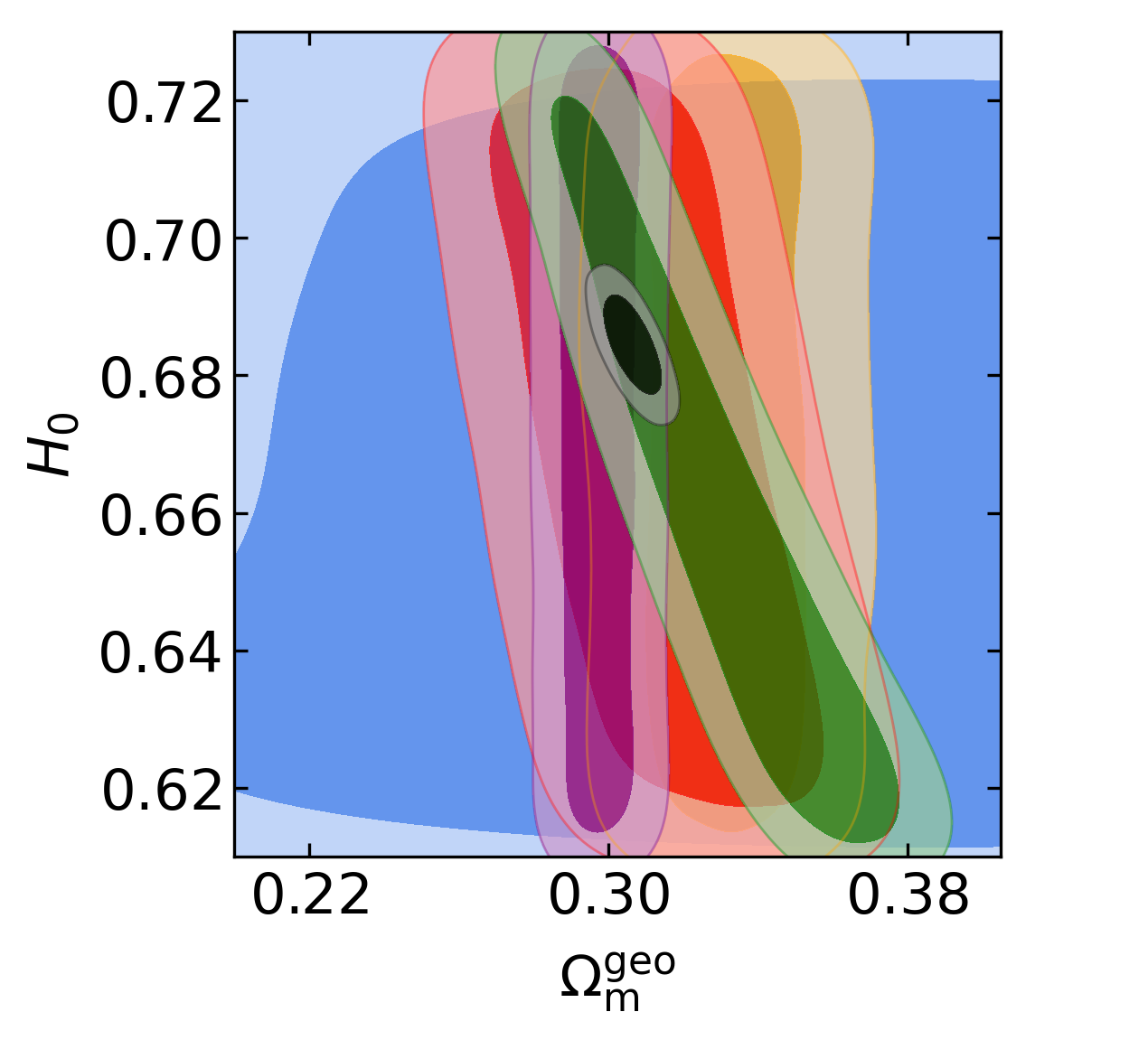}
    	\label{fig:2D-geo-h0}
	\end{subfigure}
    \hspace{-0.5cm}
    \begin{subfigure}{0.24\textwidth}
        \includegraphics[width=\linewidth]{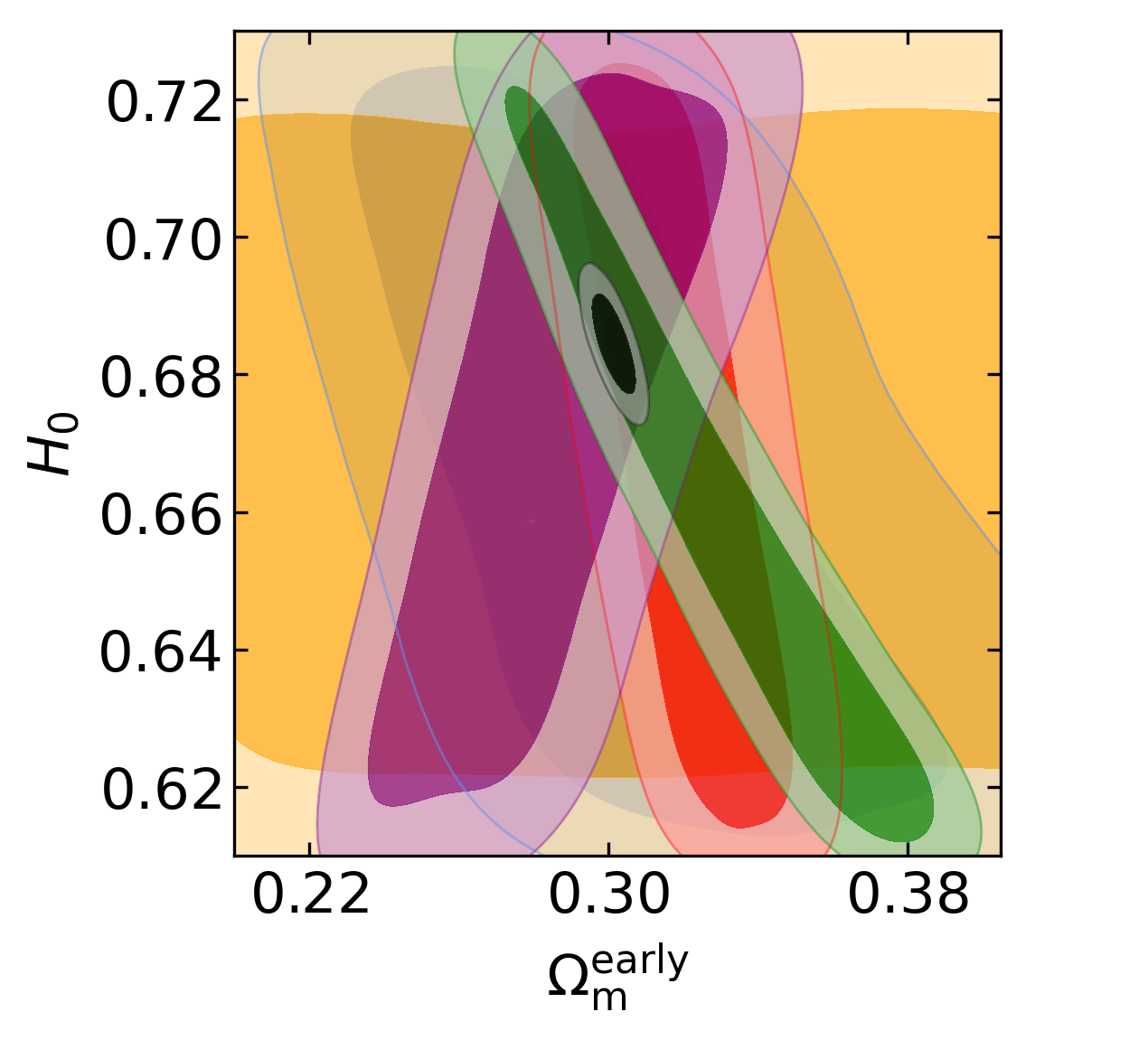}
	   \label{fig:2D-early-h0}
    \end{subfigure}
    \caption{2D posterior distribution of RSD in blue, Pantheon+ SNe Ia in orange, DES Y3 3x2pt in red, DESI DR2 BAO in purple, \textit{Planck} CMB with scale cuts in green, and all probes combined in black. Left: 2D posterior distribution for $\Om^{\rm geo}$ and $H_0$. Right: 2D posterior distribution for $\Om^{\rm early}$ and $H_0$.
    }
    \label{fig:2D-h0}
\end{figure}

First, we show single probes and their posterior distributions to infer how the joint constraints come to be. We show the 68\% and 95\% confidence intervals using the \texttt{getdist} package \citep{Lewis:2019xzd}. The correlations of $H_0$ and $\Om^{\rm geo}$ are shown in the left panel of Fig.\,\ref{fig:2D-h0} . Since \textit{Planck} constrains the geometric physical matter density $\omega_{\rm m}^{\rm geo}= \Om^{\rm geo}h^2$, this induces a negative correlation between the $H_0 = 100 h$ and the $\Om^{\rm geo}$ parameters. We did not use BBN data, and without these, DESI data do not constrain the value of $H_0$. However, they do constrain the matter density, which breaks the degeneracy of the \textit{Planck} constraint. DESI also constrains the early regime, but the main source of information is geometric. For this reason, the $\Om$ discrepancy between Pantheon+ \citep{Brout:2022vxf} and DESI \citep{DESI:2025zgx} is mostly reproduced here. The combined best-fit lies between the best-fit values of the two surveys.

Focusing on the early regime of the matter density and its contour with $H_0$ in Fig.\,\ref{fig:2D-h0} on the right panel, we see similar but not equal correlations. For the BAO data, the correlation between $H_0$ and $\Om^{\rm early}$ is positive. This comes from the sound horizon in Eq.\,(\ref{r_d_equation_desi}), taking into account the first and third terms. SNe Ia data do not measure either parameter and are only plotted for completeness here. As before, the probe complementarity leads to a tight joint constraint. DES on its own does not provide great constraints on every parameter. However, it provides different degeneracies than the other probes and is helpful when combining the different observables.

\begin{figure}
	\centering
    \begin{subfigure}{0.35\textwidth}
        \includegraphics[width=\linewidth]{legend-rsd.png}
    \end{subfigure}
    \begin{subfigure}{0.24\textwidth}
    	\includegraphics[width=\linewidth]{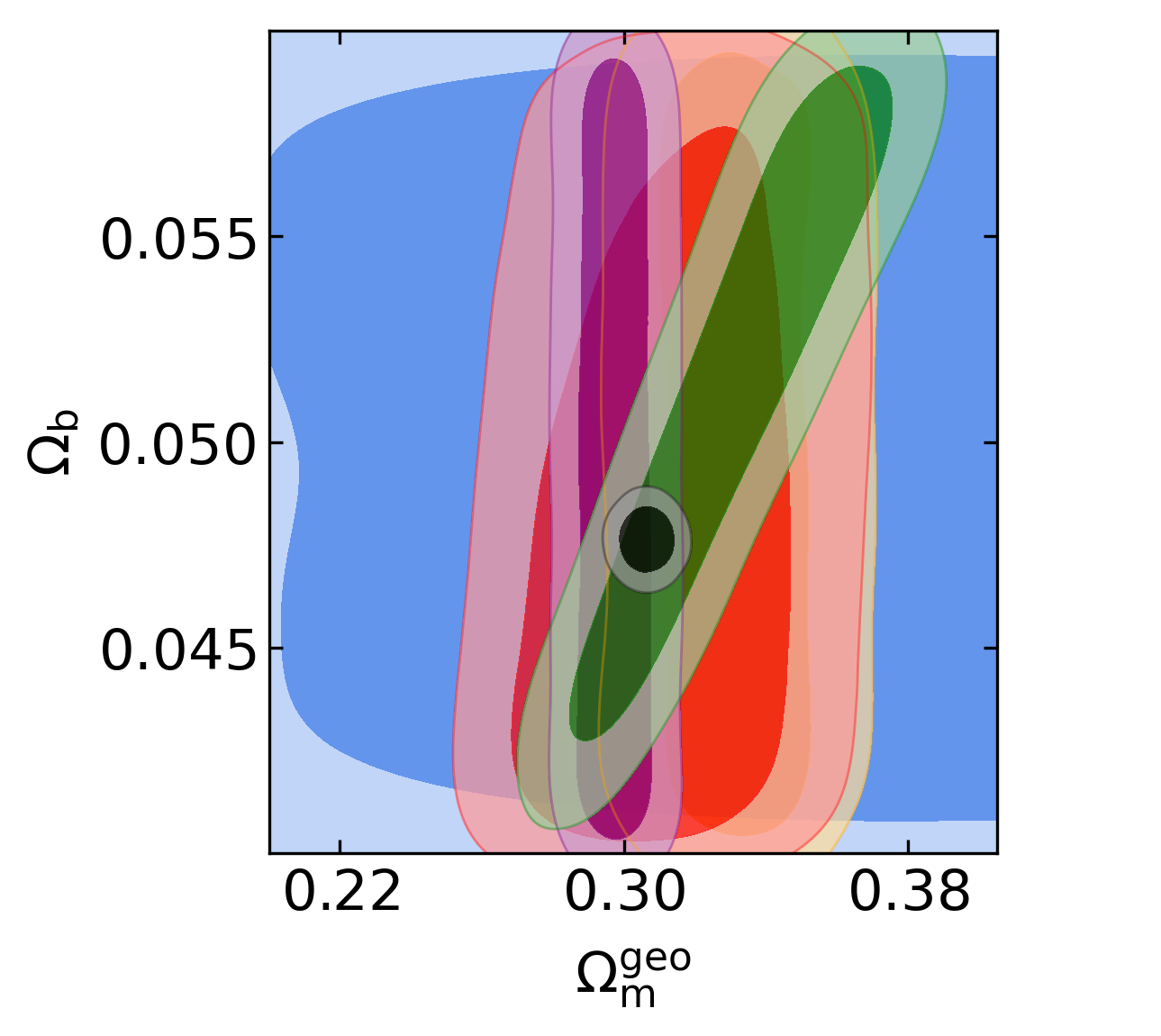}
    \end{subfigure}
    \hspace{-0.5cm}
    \begin{subfigure}{0.24\textwidth}
        \includegraphics[width=\linewidth]{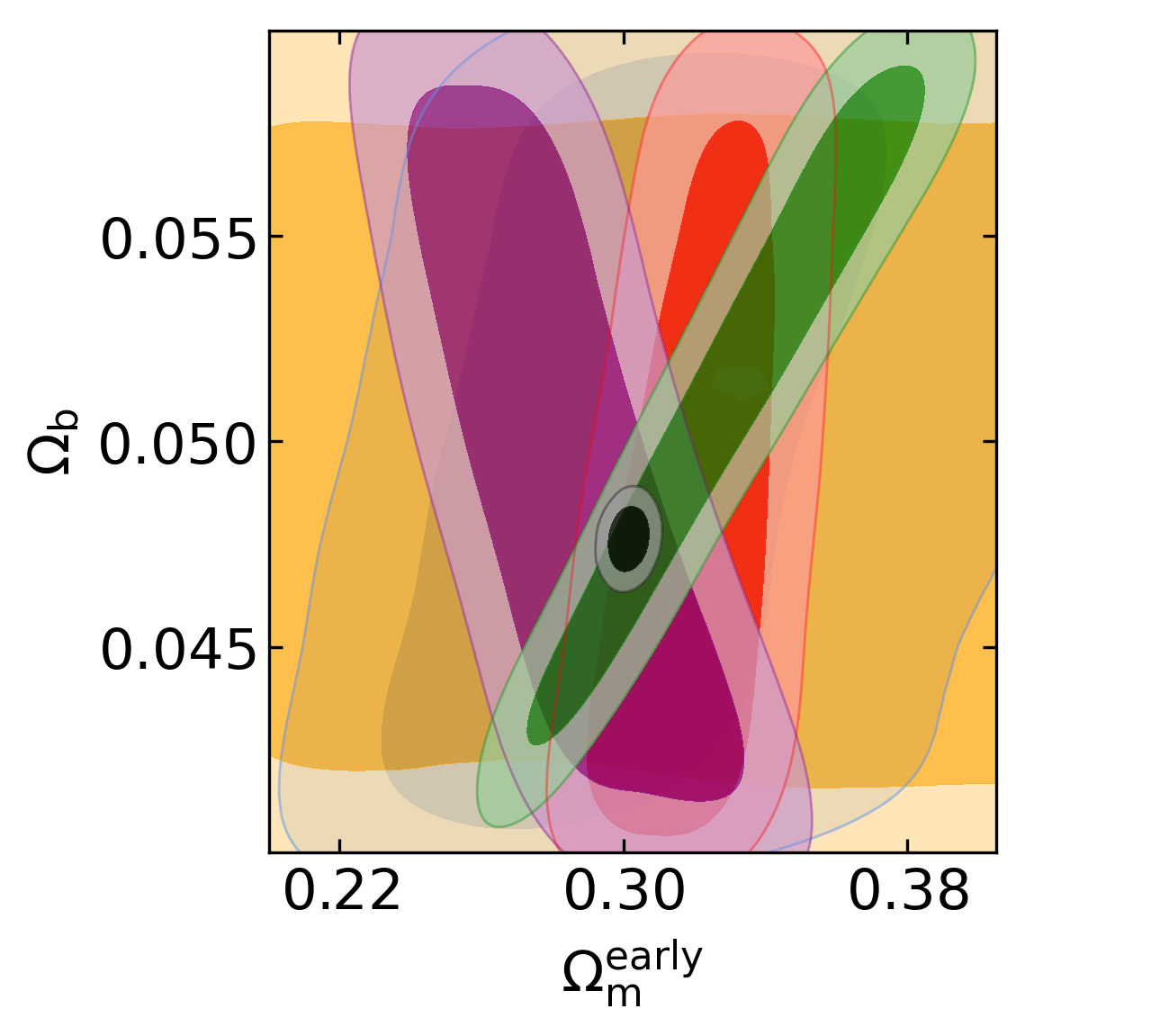}
    \end{subfigure}
    \begin{subfigure}{0.35\textwidth}
        \includegraphics[width=\linewidth]{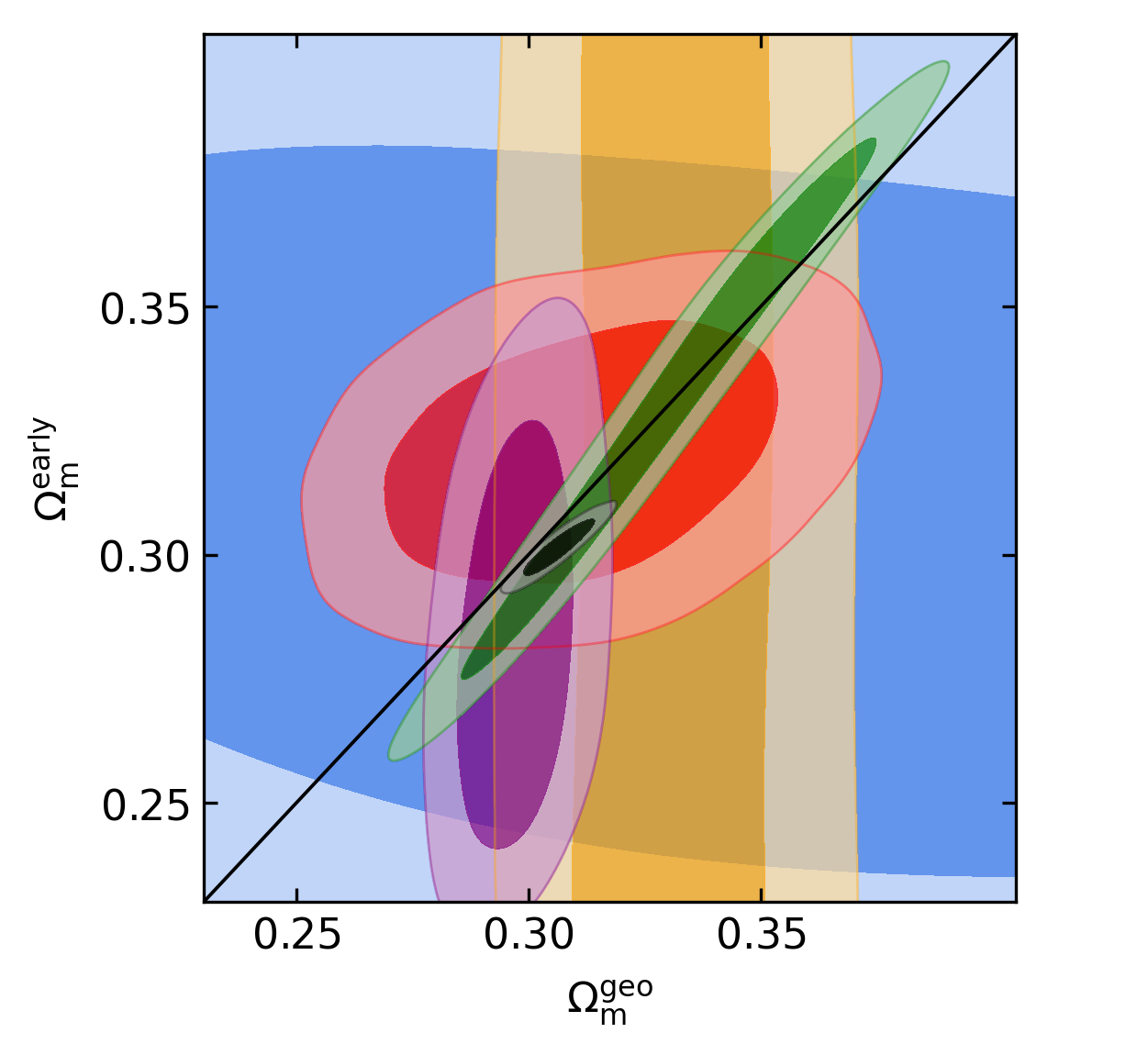}
    \end{subfigure}
    \caption{2D posterior distribution of RSD in blue, Pantheon+ SNe Ia in orange, DES Y3 3x2pt in red, DESI DR2 BAO in purple, \textit{Planck} CMB with scale cuts in green, and all probes combined in black. Top left: 2D posterior distribution for $\Om^{\rm geo}$ and $\Omega_{\rm b}$. Top right: 2D posterior distribution for $\Om^{\rm early}$ and $\Omega_{\rm b}$. Bottom: 2D posterior distribution for $\Om^{\rm geo}$ and $\Om^{\rm early}$. The black line indicates $\Om^{\rm geo}=\Om^{\rm early}$.}
    \label{fig:2D-ob}
\end{figure}

We show the 2D contour of $\Om^{\rm geo}$ and $\Omega_{\rm b}$ in the left panel of Fig.\,\ref{fig:2D-ob}. In the \textit{Planck} constraint, they are correlated, since the CMB measures both parameters and a higher baryon density naturally leads to a higher matter density. The other probes do not measure $\Ob$, but they can break the degeneracy with their constraint on $\Om^{\rm geo}$, especially DESI.

When we compare $\Om^{\rm early}$ with the baryon density $\Ob$, their correlation in \textit{Planck} is similar (see the right panel of Fig.\,\ref{fig:2D-ob}). For the BAO, we find a negative correlation because the two parameters are related to the value of the sound horizon $r_{\rm d}$. The SNe Ia sample does not  measure $\Om^{\rm early}$ since the luminosity distance only depends on $\Om^{\rm geo}$.

\begin{figure*}[h]
	\centering
    \sidecaption
	\includegraphics[
    width=12cm]{
    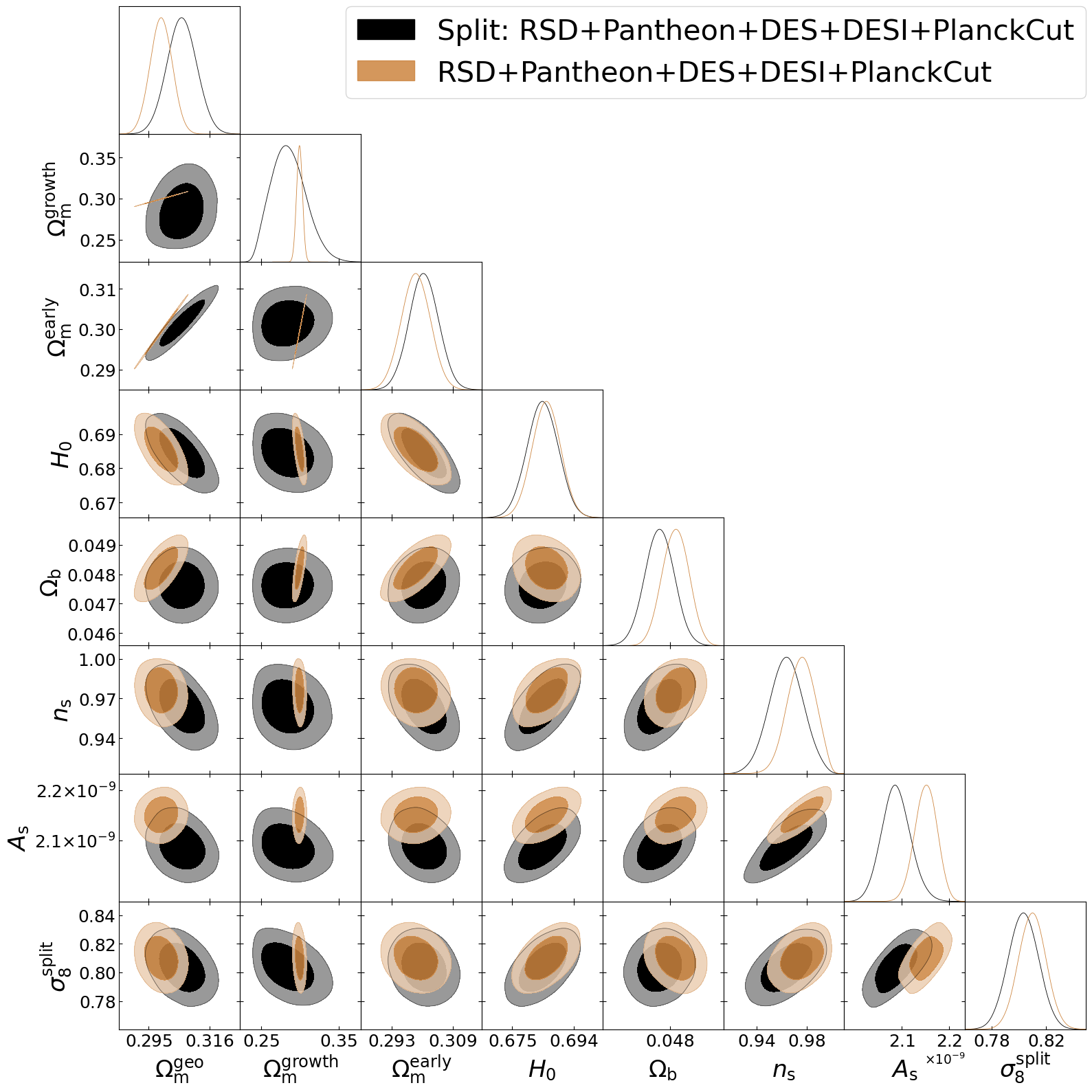}
	\caption{Posterior distributions for all probes combined (Pantheon+, DES Y3, \textit{Planck} with scale cuts (PlanckCut), DESI DR2, and RSD), 
    showing the standard likelihood in brown and the split likelihood in black.}
	\label{fig:triangle-split-nosplit}
\end{figure*}
The early and the geometry regimes are strongly correlated in the \textit{Planck} likelihood, which is shown in the bottom panel of Fig.\,\ref{fig:2D-ob}. The correlation in DESI and DES data is weaker. DES data constrain the early regime slightly better than the geometry regime because the linear matter power spectrum is based on $\Om^{\rm early}$, whereas the geometry regime enters mostly through the window functions of GC and WL. However, $\Om^{\rm geo}$ is better constrained with BAO data, which mostly give us information about cosmological distances. The combination of these probes breaks degeneracies and leads to strong constraints in both regimes.

We showed that the different probes constrain cosmological parameters and how they are used to break their respective degeneracies. We proceed to show the combined constraints and how they compare to the same data combination without using the parameter split in Fig.\,\ref{fig:triangle-split-nosplit}. The probes we used are those described in Sect.\,\ref{sec:data} combined, that is, DES Y3, DESI DR2, Pantheon+ (without SH0ES), RSD from \cite{Blanchard:2022xkk}, and \textit{Planck} 2018 without lensing and using scale cuts, as described in Sect.\,\ref{planck}. 
We tested how the split model performs compared to the standard analysis using the Akaike information criterion (AIC),

\begin{equation}
    \Delta \rm{AIC}=\Delta\chi^2+2\Delta p\, .
\end{equation}
Our split model has two additional parameters because there are three $\Om$ parameters compared to one $\Om$. The difference in the AIC between the standard non-split and the split model is $\Delta \rm{AIC}\approx 12.3$ and slightly favours the split model.

Fig.\,\ref{fig:triangle-split-nosplit} shows that $\Om^{\rm geo}$ and $\Om^{\rm early}$ are very well constrained and strongly correlated. The parameter $\Om^{\rm growth}$ is less constrained than the other two regimes. Compared to the standard analysis, $\Om^{\rm early}$ shows no significant shift. $\Om^{\rm geo}$ is shifted to a higher value, whereas $\Om^{\rm growth}$ is shifted to a lower value, but with a much larger uncertainty. The growth regime is only very slightly correlated with the other two regimes. This is partially due to the large difference in constraining power. In future work, we will consider and model CMB lensing to also consider growth with CMB data. 

The uncertainties of the cosmological parameters in the split scenario are slightly larger, which is expected since we enlarged the parameter space. Most correlations between parameters are stronger in the standard case. In the non-split case, $\Om$ is strongly correlated with $H_0$ because the CMB measures the physical density $\omega_{\rm m}=\Om h^2$. In the split case, $H_0$ is correlated with $\Omega_{\rm m}^{\rm geo}$ and $\Omega_{\rm m}^{\rm early}$, but less strongly. 

$H_0$ and $\sigma_8$ coincide with their non-split value. However, for $\Ob$, $n_{\rm s}$, and $A_{\rm s}$, the best-fit value is shifted lower. The strongest shift occurs for the scalar power spectrum amplitude, which is negatively correlated with the matter density in the geometry regime. This correlation comes from the WL probe, where either a higher $\Om^{\rm geo}$ in the window function or a higher $A_{\rm s}$ leads to a higher angular power spectrum prediction. Since the former is shifted higher, this also lowers $A_{\rm s}$.

\begin{figure}
	\centering
	\includegraphics[width=0.8\linewidth]{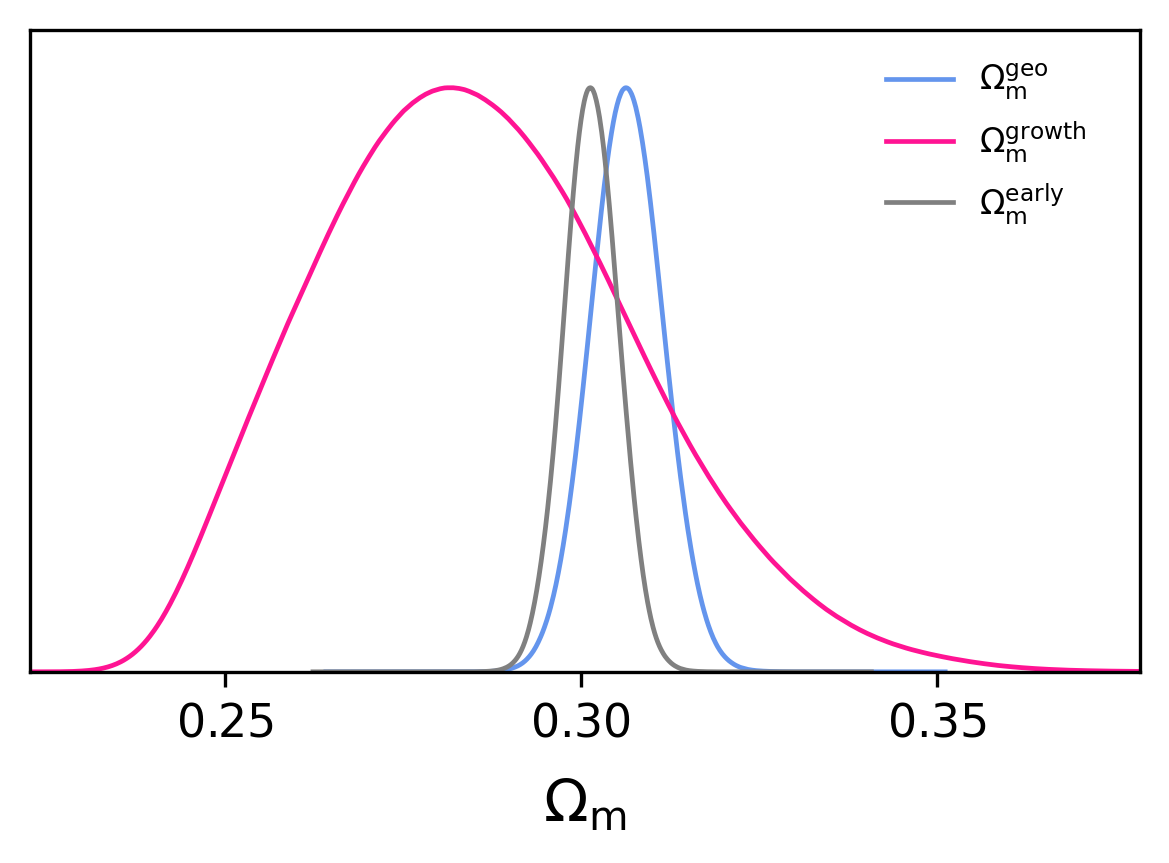}
	\caption{1D posterior distributions for all probes combined (Pantheon+, DES Y3, \textit{Planck} with scale cuts (PlanckCut), DESI DR2, and RSD) of $\Om$ in the three regimes, $\Om^{\rm geo}$ in blue, $\Om^{\rm growth}$ in pink, and $\Om^{\rm early}$ in grey.
	\label{fig:1D-omega-m}}
\end{figure}

Focusing on the matter density, we show the 1D posterior distributions in Fig.\,\ref{fig:1D-omega-m}. We find $\Om^{\rm geo} = 0.3064 \pm 0.0051 $, $\Om^{\rm growth} = 0.2855 \pm 0.0221$, and $\Om^{\rm early} = 0.3015 \pm 0.0038$. The values for geometry and growth agree within $0.9\sigma$. The difference between the early and the growth regimes is  $0.7\sigma$. The values for the geometry and early regimes are close, but their uncertainties are also much smaller. This leads to a difference of $0.8\sigma$. All in all, these three values are compatible with each other within $1\sigma$.

\begin{figure}
	\centering
	\includegraphics[width=\linewidth]{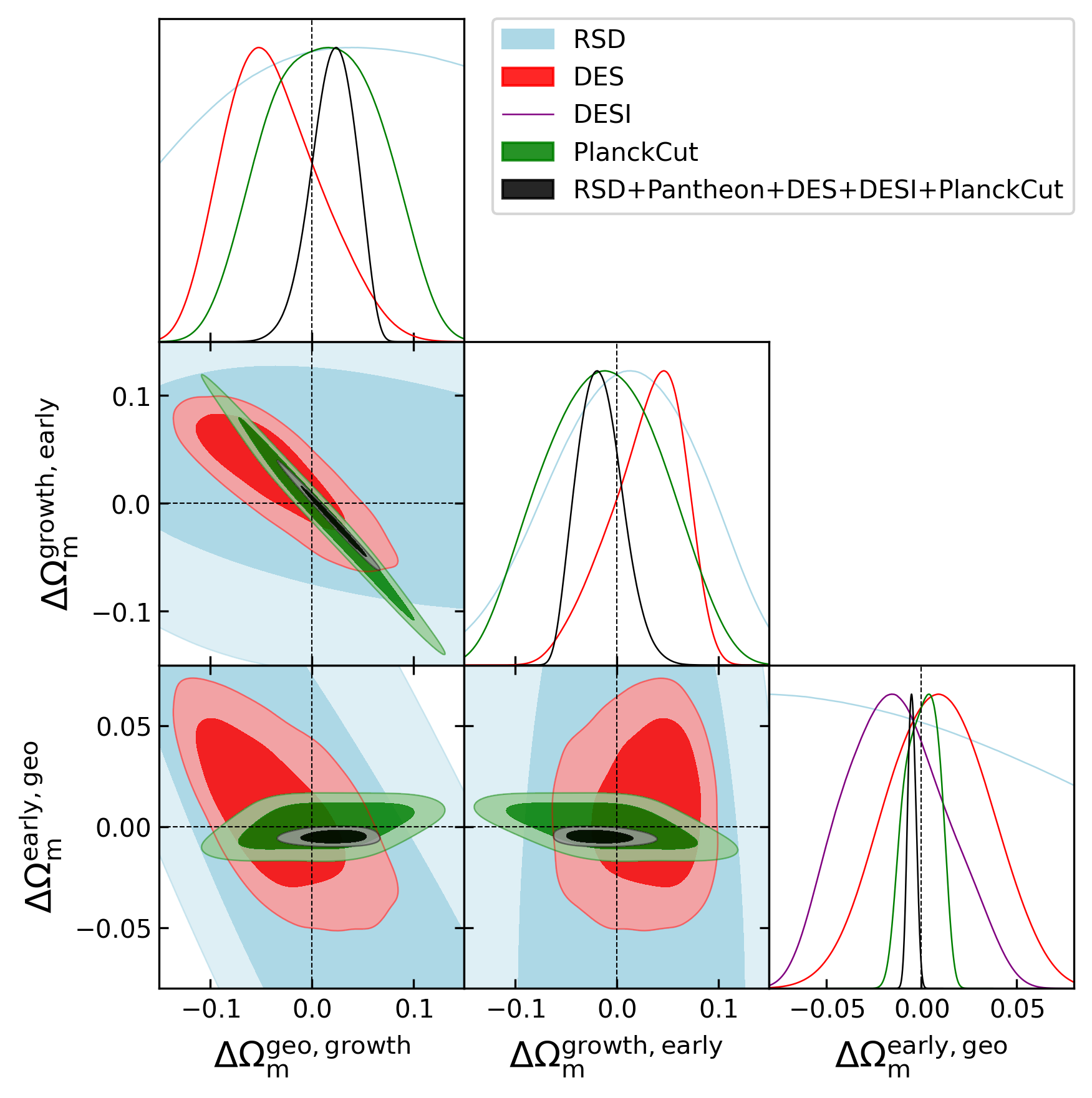}
	\caption{Posterior distribution of the difference between the $\Om$ parameters in two regimes. We show the contours of the different probes, DES Y3 3x2pt in red, DESI DR2 BAO in purple, \textit{Planck} CMB with scale cuts in green, and all probes combined (including Pantheon+ SNe Ia) in black.
	\label{fig:triangle-single-probes-delta}}
\end{figure}

We note that within the datasets, there is a discrepancy of $2\sigma$ between the DESI DR2 value at $\Om=0.2975 \pm 0.0086$ \citep{DESI:2025zgx} and Pantheon+ at $\Om=0.334\pm0.018$ \citep{Brout:2022vxf}. The \textit{Planck} value of $\Om=0.3153 \pm 0.0073$ lies in the middle, with an error comparable to the one of DESI DR2, leading to a discrepancy of $2\sigma$ as well. Our constraints do not directly mirror this discrepancy, but extract different information from these same datasets.

To study the discrepancies between two regimes in more detail, we show the posteriors of the differences between two $\Om$, using parameters that quantify the difference $\Delta\Om^{i,j} = \Om^{i}-\Om^{j}$. These posterior distributions are shown in Fig.\,\ref{fig:triangle-single-probes-delta}. We note that only the 3x2pt, the CMB, and the RSD probes constrain all three regimes, and thus, only their 2D contours can be shown in this figure. The BAO probe constrains the early and the geometry regime and is shown in the corresponding 1D contour.

The SNe Ia data contribute to the joint constraint, but cannot solely constrain the difference of parameters because they only probe the geometric regime. The posterior for the scale-cut {\it Planck} likelihood is slightly affected by the priors on the growth regime that stem from \texttt{EuclidEmulator2} ($0.24 < \Om^{\rm growth} < 0.4$). However, this does not affect the combined posterior.

The combined contours of the differences $\Delta\Om^{\rm geo, growth}$ and $\Delta\Om^{\rm early, growth}$ peak close to 0, within 1$\sigma$. However, the difference between the early and the geometry regimes, $\Delta\Om^{\rm geo, early}$, is different from 0 by 2$\sigma$. 

Figs. \ref{fig:triangle-split-nosplit} and \ref{fig:triangle-single-probes-delta} show that the difference between the two parameters can be much better constrained than their absolute value because they are correlated (similar to the reasoning for the definition of $S_8$). Therefore, the uncertainty on $\Delta\Om^{\rm geo, early}$ is only 0.002. This leads to the discrepancy in the $\Delta\Om^{\rm geo, early}$ parameter, even though the $\Om$ values themselves are compatible within $1\sigma$.

\section{Conclusion}\label{sec:conclusions}
In light of the tensions that appear in $\Lambda$CDM and the multitude of proposed extensions, it is important to subject the standard model to consistency tests. This diagnoses potential problems in $\Lambda$CDM, especially when tested across multiple observables, as we did here.

We proposed a new parameter split that considered three cosmological regimes: geometry, structure growth, and the early Universe. We considered multiple cosmological probes such that their combination was sensitive to all the regimes in different ways. This ensured that our combined constraints were not overly affected by a single survey and its potential systematics. We separated the matter density parameter $\Om$ in the likelihoods of the 3x2pt, CMB, BAO, SNe Ia, and RSD probes into these regimes.

For photometric 3x2pt data, we used a formalism that captured the sensitivity of the observable to all three $\Om$ parameters. These data are sensitive to the early Universe through the linear matter power spectrum, to geometry through the window functions, and also to structure formation through non-linearities and the growth factor. For the CMB likelihood, we used scale cuts and did not add CMB lensing data to exclude probes that are affected by structure growth. The early regime is constrained by CMB data, but so is the geometry regime, which governs our distance to the last-scattering surface. The BAO data are sensitive to the same two regimes. They are mostly governed by geometry, since the observables are distances. However, the early regime comes into play through the sound horizon at drag redshift. Furthermore, we considered SNe Ia, which only give us information about the geometry because they measure the luminosity distance.  RSD data are sensitive to the early regime through the EBS quantities affecting $\sigma_8$ and to the growth regime through the rescaling of $\sigma_8$ and the growth factor $f(z)$. They are also sensitive to geometry to a lesser extent through geometric factors in the differential equation for $f(z)$.

We tested this parameter split on each probe separately to determine where the constraints originate. To do this, we showed how the split $\Om$ parameters are correlated with each other and with the $H_0$ and $\Ob$ parameters for the different probes we used. In the BAO data, there is no significant correlation between $\Om$ in the geometry versus the early regime. For the 3x2pt data, we found a small correlation. However, our scale-cut $\textit{Planck}$ likelihood shows a very strong correlation between these parameters. This correlation carries over to the joint constraint.

We further compared our split results to the same data combination in a standard non-split setup to show how the parameters shift. The matter density inferred from early-universe information, $\Om^{\rm early}$, is largely the same. The geometric matter density $\Om^{\rm geo}$ is shifted to a slightly higher best-fit value, whereas $\Om^{\rm growth}$ is shifted to a lower value, but with a larger uncertainty. 

As in the literature \citep[see, e.g.,][]{DES:2020iqt, Ruiz-Zapatero:2021rzl, Zhong:2023how}, we found that these cosmological observables provide tighter constraints on geometry than on growth. Additionally, we found that we were able to constrain the early and the geometric regime at the same time, whereas the larger error on the growth regime persisted.

For the surveys that constrain multiple regimes and for the combined datasets, we showed the posteriors for the parameters $\Delta \Om$ that quantify the differences between two regimes. Whereas the absolute $\Om$ values agree, the $\Delta\Om^{\rm geo, early}$ value shows a $2\sigma$ discrepancy with 0 because the difference can be better constrained than the parameters on their own.

In conclusion, we have found that all three regimes are compatible. To improve the constraints on the growth regime, we plan to include more observables that contain information about structure formation. In particular, we will consider clusters of galaxies, that is, number counts and cluster clustering, on the one hand and CMB lensing, which we have explicitly excluded here, on the other hand.

\begin{acknowledgements}
We would like to thank Ziad Sakr, Matteo Martinelli, Sefa Pamuk, and Savvas Nesseris for useful discussions and Peter Taylor for the help regarding the use of \texttt{CombineHarvesterFlow}.
IT has been supported by the Ramon y Cajal fellowship (RYC2023-045531-I) funded by the State Research Agency of the Spanish Ministerio de Ciencia, Innovaci\'on y Universidades, MICIU/AEI/10.13039/501100011033, and Social European Funds plus (FSE+). IT also acknowledges support from the same ministry, via projects PID2022-141079NB, PID2022-138896NB; the European Research Executive Agency HORIZON-MSCA-2021-SE-01 Research and Innovation programme under the Marie Sk\l odowska-Curie grant agreement number 101086388 (LACEGAL) and the programme Unidad de Excelencia Mar\'{\i}a de Maeztu, project CEX2020-001058-M.
\end{acknowledgements}

\bibliography{omega-m-3-split.bib}

@article{Andrade:2021njl,
    author = "Andrade, Uendert and Anbajagane, Dhayaa and von Marttens, Rodrigo and Huterer, Dragan and Alcaniz, Jailson",
    title = "{A test of the standard cosmological model with geometry and growth}",
    eprint = "2107.07538",
    archivePrefix = "arXiv",
    primaryClass = "astro-ph.CO",
    doi = "10.1088/1475-7516/2021/11/014",
    journal = "JCAP",
    volume = "11",
    pages = "014",
    year = "2021"
}

@article{Bertolez-Martinez:2024wez,
    author = "Bert{\'o}lez-Mart{\'\i}nez, Toni and Esteban, Ivan and Hajjar, Rasmi and Mena, Olga and Salvado, Jordi",
    title = "{Origin of cosmological neutrino mass bounds: background versus perturbations}",
    eprint = "2411.14524",
    archivePrefix = "arXiv",
    primaryClass = "astro-ph.CO",
    doi = "10.1088/1475-7516/2025/06/058",
    journal = "JCAP",
    volume = "06",
    pages = "058",
    year = "2025"
}

@article{Abate:2008au,
    author = "Abate, Alexandra and Lahav, Ofer",
    title = "{The Three Faces of Omega{\_}m: Testing Gravity with Low and High Redshift SN Ia Surveys}",
    eprint = "0805.3160",
    archivePrefix = "arXiv",
    primaryClass = "astro-ph",
    doi = "10.1111/j.1745-3933.2008.00519.x",
    journal = "Mon. Not. Roy. Astron. Soc.",
    volume = "389",
    pages = "47",
    year = "2008"
}

@article{Bernal:2015zom,
    author = "Bernal, Jos{\'e} Luis and Verde, Licia and Cuesta, Antonio J.",
    title = "{Parameter splitting in dark energy: is dark energy the same in the background and in the cosmic structures?}",
    eprint = "1511.03049",
    archivePrefix = "arXiv",
    primaryClass = "astro-ph.CO",
    doi = "10.1088/1475-7516/2016/02/059",
    journal = "JCAP",
    volume = "02",
    pages = "059",
    year = "2016"
}

@article{Zhong:2024xuk,
    author = "Zhong, Kunhao and Saraivanov, Evan and Caputi, James and Miranda, Vivian and Boruah, Supranta S. and Eifler, Tim and Krause, Elisabeth",
    title = "{Attention-based neural network emulators for multiprobe data vectors. I. Forecasting the growth-geometry split}",
    eprint = "2402.17716",
    archivePrefix = "arXiv",
    primaryClass = "astro-ph.CO",
    doi = "10.1103/PhysRevD.111.123519",
    journal = "Phys. Rev. D",
    volume = "111",
    number = "12",
    pages = "123519",
    year = "2025"
}

@article{LSST:2008ijt,
    author = "Ivezi{\'c}, {\v{Z}}eljko and others",
    collaboration = "LSST",
    title = "{LSST: from Science Drivers to Reference Design and Anticipated Data Products}",
    eprint = "0805.2366",
    archivePrefix = "arXiv",
    primaryClass = "astro-ph",
    reportNumber = "SLAC-PUB-16076",
    doi = "10.3847/1538-4357/ab042c",
    journal = "Astrophys. J.",
    volume = "873",
    number = "2",
    pages = "111",
    year = "2019"
}

@article{DES:2020ypx,
    author = "Friedrich, O. and others",
    collaboration = "DES",
    title = "{Dark Energy Survey year 3 results: covariance modelling and its impact on parameter estimation and quality of fit}",
    eprint = "2012.08568",
    archivePrefix = "arXiv",
    primaryClass = "astro-ph.CO",
    reportNumber = "FERMILAB-PUB-20-663-E-SCD-V, DES-2019-0466",
    doi = "10.1093/mnras/stab2384",
    journal = "Mon. Not. Roy. Astron. Soc.",
    volume = "508",
    number = "3",
    pages = "3125--3165",
    year = "2021"
}

@book{dodelson_modern_2003,
	address = {San Diego, Calif},
	title = {Modern cosmology},
	isbn = {978-0-12-219141-1},
	publisher = {Academic Press},
	author = {Dodelson, Scott},
	year = {2003},
}

@article{Taylor2024CombineHarvesterFlow,
	author = {Taylor, Peter L. and Cuceu, Andrei and To, Chun-Hao and Zaborowski, Erik A.},
	journal = {The Open Journal of Astrophysics},
	doi = {10.33232/001c.124495},
	year = {2024},
	month = {oct 8},
	publisher = {Maynooth Academic Publishing},
	title = {CombineHarvesterFlow: Joint {Probe} {Analysis} {Made} {Easy} with {Normalizing} {Flows}},
	volume = {7},
}

@article{Lange:2023ydq,
    author = "Lange, Johannes U.",
    title = "{nautilus: boosting Bayesian importance nested sampling with deep learning}",
    eprint = "2306.16923",
    archivePrefix = "arXiv",
    primaryClass = "astro-ph.IM",
    doi = "10.1093/mnras/stad2441",
    journal = "Mon. Not. Roy. Astron. Soc.",
    volume = "525",
    number = "2",
    pages = "3181--3194",
    year = "2023"
}

@article{Euclid:2020rfv,
    author = "Knabenhans, M. and others",
    collaboration = "Euclid",
    title = "{Euclid preparation: IX. EuclidEmulator2 {\textendash} power spectrum emulation with massive neutrinos and self-consistent dark energy perturbations}",
    eprint = "2010.11288",
    archivePrefix = "arXiv",
    primaryClass = "astro-ph.CO",
    doi = "10.1093/mnras/stab1366",
    journal = "Mon. Not. Roy. Astron. Soc.",
    volume = "505",
    number = "2",
    pages = "2840--2869",
    year = "2021"
}

@article{Percival:2008sh,
    author = "Percival, Will J and White, Martin",
    title = "{Testing cosmological structure formation using redshift-space distortions}",
    eprint = "0808.0003",
    archivePrefix = "arXiv",
    primaryClass = "astro-ph",
    doi = "10.1111/j.1365-2966.2008.14211.x",
    journal = "Mon. Not. Roy. Astron. Soc.",
    volume = "393",
    pages = "297",
    year = "2009"
}

@article{DES:2021zxv,
    author = "Pandey, S. and others",
    collaboration = "DES",
    title = "{Dark Energy Survey year 3 results: Constraints on cosmological parameters and galaxy-bias models from galaxy clustering and galaxy-galaxy lensing using the redMaGiC sample}",
    eprint = "2105.13545",
    archivePrefix = "arXiv",
    primaryClass = "astro-ph.CO",
    reportNumber = "FERMILAB-PUB-21-249-SCD-T",
    doi = "10.1103/PhysRevD.106.043520",
    journal = "Phys. Rev. D",
    volume = "106",
    number = "4",
    pages = "043520",
    year = "2022"
}

@article{Hernandez:2016xci,
    author = "Hern{\'a}ndez, Oscar F.",
    title = "{Neutrino Masses, Scale-Dependent Growth, and Redshift-Space Distortions}",
    eprint = "1608.08298",
    archivePrefix = "arXiv",
    primaryClass = "astro-ph.CO",
    doi = "10.1088/1475-7516/2017/06/018",
    journal = "JCAP",
    volume = "06",
    pages = "018",
    year = "2017"
}

@article{Kilbinger:2017lvu,
    author = "Kilbinger, Martin and others",
    title = "{Precision calculations of the cosmic shear power spectrum projection}",
    eprint = "1702.05301",
    archivePrefix = "arXiv",
    primaryClass = "astro-ph.CO",
    doi = "10.1093/mnras/stx2082",
    journal = "Mon. Not. Roy. Astron. Soc.",
    volume = "472",
    number = "2",
    pages = "2126--2141",
    year = "2017"
}

@article{Brout:2022vxf,
    author = "Brout, Dillon and others",
    title = "{The Pantheon+ Analysis: Cosmological Constraints}",
    eprint = "2202.04077",
    archivePrefix = "arXiv",
    primaryClass = "astro-ph.CO",
    doi = "10.3847/1538-4357/ac8e04",
    journal = "Astrophys. J.",
    volume = "938",
    number = "2",
    pages = "110",
    year = "2022"
}

@article{Scolnic:2021amr,
    author = "Scolnic, Dan and others",
    title = "{The Pantheon+ Analysis: The Full Data Set and Light-curve Release}",
    eprint = "2112.03863",
    archivePrefix = "arXiv",
    primaryClass = "astro-ph.CO",
    doi = "10.3847/1538-4357/ac8b7a",
    journal = "Astrophys. J.",
    volume = "938",
    number = "2",
    pages = "113",
    year = "2022"
}

@article{Tripp:1997wt,
    author = "Tripp, Robert",
    title = "{A Two-parameter luminosity correction for type Ia supernovae}",
    reportNumber = "LBL-40857, LBNL-40857",
    journal = "Astron. Astrophys.",
    volume = "331",
    pages = "815--820",
    year = "1998"
}

@article{BOSS:2016wmc,
    author = "Alam, Shadab and others",
    collaboration = "BOSS",
    title = "{The clustering of galaxies in the completed SDSS-III Baryon Oscillation Spectroscopic Survey: cosmological analysis of the DR12 galaxy sample}",
    eprint = "1607.03155",
    archivePrefix = "arXiv",
    primaryClass = "astro-ph.CO",
    doi = "10.1093/mnras/stx721",
    journal = "Mon. Not. Roy. Astron. Soc.",
    volume = "470",
    number = "3",
    pages = "2617--2652",
    year = "2017"
}

@article{PhysRevD.100.043504,
  title = {Quantifying tensions in cosmological parameters: Interpreting the DES evidence ratio},
  author = {Handley, Will and Lemos, Pablo},
  journal = {Phys. Rev. D},
  volume = {100},
  issue = {4},
  pages = {043504},
  numpages = {15},
  year = {2019},
  month = {Aug},
  publisher = {American Physical Society},
  doi = {10.1103/PhysRevD.100.043504},
  url = {https://link.aps.org/doi/10.1103/PhysRevD.100.043504}
}

@article{Planck:2015bue,
    author = "Ade, P. A. R. and others",
    collaboration = "Planck",
    title = "{Planck 2015 results. XIV. Dark energy and modified gravity}",
    eprint = "1502.01590",
    archivePrefix = "arXiv",
    primaryClass = "astro-ph.CO",
    doi = "10.1051/0004-6361/201525814",
    journal = "Astron. Astrophys.",
    volume = "594",
    pages = "A14",
    year = "2016"
}

@article{Cooke:2016rky,
    author = "Cooke, Ryan J. and Pettini, Max and Nollett, Kenneth M. and Jorgenson, Regina",
    title = "{The primordial deuterium abundance of the most metal-poor damped Ly$\alpha$ system}",
    eprint = "1607.03900",
    archivePrefix = "arXiv",
    primaryClass = "astro-ph.CO",
    doi = "10.3847/0004-637X/830/2/148",
    journal = "Astrophys. J.",
    volume = "830",
    number = "2",
    pages = "148",
    year = "2016"
}

@article{Pan-STARRS1:2017jku,
    author = "Scolnic, D. M. and others",
    collaboration = "Pan-STARRS1",
    title = "{The Complete Light-curve Sample of Spectroscopically Confirmed SNe Ia from Pan-STARRS1 and Cosmological Constraints from the Combined Pantheon Sample}",
    eprint = "1710.00845",
    archivePrefix = "arXiv",
    primaryClass = "astro-ph.CO",
    doi = "10.3847/1538-4357/aab9bb",
    journal = "Astrophys. J.",
    volume = "859",
    number = "2",
    pages = "101",
    year = "2018"
}

@article{Colgain:2024mtg,
    author = "Colg{\'a}in, Eoin {\'O}. and Sheikh-Jabbari, M. M.",
    title = "{DESI and SNe: Dynamical Dark Energy, $\Omega_m$ Tension or Systematics?}",
    eprint = "2412.12905",
    archivePrefix = "arXiv",
    primaryClass = "astro-ph.CO",
    doi = "10.1093/mnrasl/slaf042",
    journal = "Mon. Not. Roy. Astron. Soc.",
    volume = "542",
    number = "1",
    pages = "L24--L30",
    year = "2025"
}

@article{Weinberg:1988cp,
    author = "Weinberg, Steven",
    editor = "Hsu, Jong-Ping and Fine, D.",
    title = "{The Cosmological Constant Problem}",
    reportNumber = "UTTG-12-88",
    doi = "10.1103/RevModPhys.61.1",
    journal = "Rev. Mod. Phys.",
    volume = "61",
    pages = "1--23",
    year = "1989"
}

@article{KiDS:2020suj,
    author = "Asgari, Marika and others",
    collaboration = "KiDS",
    title = "{KiDS-1000 Cosmology: Cosmic shear constraints and comparison between two point statistics}",
    eprint = "2007.15633",
    archivePrefix = "arXiv",
    primaryClass = "astro-ph.CO",
    doi = "10.1051/0004-6361/202039070",
    journal = "Astron. Astrophys.",
    volume = "645",
    pages = "A104",
    year = "2021"
}

@book{Weinberg:2008zzc,
    author = "Weinberg, Steven",
    title = "{Cosmology}",
	publisher = {Oxford university press},
    isbn = "978-0-19-852682-7",
    year = "2008"
}

@article{Hanson:2009kr,
    author = "Hanson, Duncan and Challinor, Anthony and Lewis, Antony",
    title = "{Weak lensing of the CMB}",
    eprint = "0911.0612",
    archivePrefix = "arXiv",
    primaryClass = "astro-ph.CO",
    doi = "10.1007/s10714-010-1036-y",
    journal = "Gen. Rel. Grav.",
    volume = "42",
    pages = "2197--2218",
    year = "2010"
}

@article{Zhang:2003ii,
    author = "Zhang, Jun and Hui, Lam and Stebbins, Albert",
    title = "{Isolating Geometry in Weak Lensing Measurements}",
    eprint = "astro-ph/0312348",
    archivePrefix = "arXiv",
    reportNumber = "FERMILAB-PUB-03-419-A",
    doi = "10.1086/497676",
    journal = "Astrophys. J.",
    volume = "635",
    pages = "806--820",
    year = "2005"
}

@article{Maus:2026wsb,
    author = "Maus, Mark and Baleato Lizancos, Ant{\'o}n and White, Martin and de Mattia, Arnaud and Chen, Shi-Fan",
    title = "{An analytic approximation to the covariance between pre- and post-reconstruction galaxy two-point statistics}",
    eprint = "2602.12343",
    archivePrefix = "arXiv",
    primaryClass = "astro-ph.CO",
    doi = "10.1088/1475-7516/2026/05/078",
    journal = "JCAP",
    volume = "05",
    pages = "078",
    year = "2026"
}

@article{Brieden:2022heh,
    author = "Brieden, Samuel and Gil-Mar\'\i{}n, H\'ector and Verde, Licia",
    title = "{A tale of two (or more) h's}",
    eprint = "2212.04522",
    archivePrefix = "arXiv",
    primaryClass = "astro-ph.CO",
    doi = "10.1088/1475-7516/2023/04/023",
    journal = "JCAP",
    volume = "04",
    pages = "023",
    year = "2023"
}

@article{Kaiser:1987qv,
    author = "Kaiser, N.",
    title = "{Clustering in real space and in redshift space}",
    doi = "10.1093/mnras/227.1.1",
    journal = "Mon. Not. Roy. Astron. Soc.",
    volume = "227",
    pages = "1--27",
    year = "1987"
}

@article{10.1093/mnras/stv1436,
    author = {Liske, J. and Baldry, I. K. and Driver, S. P. and Tuffs, R. J. and Alpaslan, M. and Andrae, E. and Brough, S. and Cluver, M. E. and Grootes, M. W. and Gunawardhana, M. L. P. and Kelvin, L. S. and Loveday, J. and Robotham, A. S. G. and Taylor, E. N. and Bamford, S. P. and Bland-Hawthorn, J. and Brown, M. J. I. and Drinkwater, M. J. and Hopkins, A. M. and Meyer, M. J. and Norberg, P. and Peacock, J. A. and Agius, N. K. and Andrews, S. K. and Bauer, A. E. and Ching, J. H. Y. and Colless, M. and Conselice, C. J. and Croom, S. M. and Davies, L. J. M. and De Propris, R. and Dunne, L. and Eardley, E. M. and Ellis, S. and Foster, C. and Frenk, C. S. and Häußler, B. and Holwerda, B. W. and Howlett, C. and Ibarra, H. and Jarvis, M. J. and Jones, D. H. and Kafle, P. R. and Lacey, C. G. and Lange, R. and Lara-López, M. A. and López-Sánchez, Á. R. and Maddox, S. and Madore, B. F. and McNaught-Roberts, T. and Moffett, A. J. and Nichol, R. C. and Owers, M. S. and Palamara, D. and Penny, S. J. and Phillipps, S. and Pimbblet,  K. A. and Popescu, C. C. and Prescott, M. and Proctor, R. and Sadler, E. M. and Sansom, A. E. and Seibert, M. and Sharp, R. and Sutherland, W. and Vázquez-Mata, J. A. and van Kampen, E. and Wilkins, S. M. and Williams, R. and Wright, A. H.},
    title = {Galaxy And Mass Assembly (GAMA): end of survey report and data release 2},
    journal = {Mon. Not. Roy. Astron. Soc.},
    volume = {452},
    number = {2},
    pages = {2087-2126},
    year = {2015},
    month = {09},
    issn = {0035-8711},
    doi = {10.1093/mnras/stv1436},
    url = {https://doi.org/10.1093/mnras/stv1436},
    eprint = {https://academic.oup.com/mnras/article-pdf/452/2/2087/18508439/stv1436.pdf},
}

@article{Lange:2021zre,
    author = "Lange, Johannes U. and Hearin, Andrew P. and Leauthaud, Alexie and van den Bosch, Frank C. and Guo, Hong and DeRose, Joseph",
    title = "{Five per{\,}cent measurements of the growth rate from simulation-based modelling of redshift-space clustering in BOSS LOWZ}",
    eprint = "2101.12261",
    archivePrefix = "arXiv",
    primaryClass = "astro-ph.CO",
    doi = "10.1093/mnras/stab3111",
    journal = "Mon. Not. Roy. Astron. Soc.",
    volume = "509",
    number = "2",
    pages = "1779--1804",
    year = "2021"
}

@article{eBOSS:2018abz,
    author = "Gil-Mar{\'\i}n, H{\'e}ctor and others",
    collaboration = "eBOSS",
    title = "{The clustering of the SDSS-IV extended Baryon Oscillation Spectroscopic Survey DR14 quasar sample: structure growth rate measurement from the anisotropic quasar power spectrum in the redshift range $0.8 < z < 2.2$}",
    eprint = "1801.02689",
    archivePrefix = "arXiv",
    primaryClass = "astro-ph.CO",
    doi = "10.1093/mnras/sty453",
    journal = "Mon. Not. Roy. Astron. Soc.",
    volume = "477",
    number = "2",
    pages = "1604--1638",
    year = "2018"
}

@article{Blake:2011rj,
    author = "Blake, Chris and others",
    title = "{The WiggleZ Dark Energy Survey: the growth rate of cosmic structure since redshift z=0.9}",
    eprint = "1104.2948",
    archivePrefix = "arXiv",
    primaryClass = "astro-ph.CO",
    doi = "10.1111/j.1365-2966.2011.18903.x",
    journal = "Mon. Not. Roy. Astron. Soc.",
    volume = "415",
    pages = "2876",
    year = "2011"
}

@article{Okumura:2015lvp,
    author = "Okumura, Teppei and others",
    title = "{The Subaru FMOS galaxy redshift survey (FastSound). IV. New constraint on gravity theory from redshift space distortions at $z\sim 1.4$}",
    eprint = "1511.08083",
    archivePrefix = "arXiv",
    primaryClass = "astro-ph.CO",
    doi = "10.1093/pasj/psw029",
    journal = "Publ. Astron. Soc. Jap.",
    volume = "68",
    number = "3",
    pages = "38",
    year = "2016"
}

@article{delaTorre:2016rxm,
    author = "de la Torre, S. and others",
    title = "{The VIMOS Public Extragalactic Redshift Survey (VIPERS). Gravity test from the combination of redshift-space distortions and galaxy-galaxy lensing at $0.5 < z < 1.2$}",
    eprint = "1612.05647",
    archivePrefix = "arXiv",
    primaryClass = "astro-ph.CO",
    doi = "10.1051/0004-6361/201630276",
    journal = "Astron. Astrophys.",
    volume = "608",
    pages = "A44",
    year = "2017"
}

@article{Koussour:2023ulc,
    author = "Koussour, M. and Myrzakulov, N. and Alfedeel, Alnadhief H. A. and Abebe, Amare",
    title = "{Constraining the cosmological model of modified f(Q) gravity: Phantom dark energy and observational insights}",
    eprint = "2310.15067",
    archivePrefix = "arXiv",
    primaryClass = "astro-ph.CO",
    doi = "10.1093/ptep/ptad133",
    journal = "PTEP",
    volume = "2023",
    number = "11",
    pages = "113E01",
    year = "2023"
}

@article{Kunz:2006wc,
    author = "Kunz, Martin and Sapone, Domenico",
    title = "{Crossing the Phantom Divide}",
    eprint = "astro-ph/0609040",
    archivePrefix = "arXiv",
    doi = "10.1103/PhysRevD.74.123503",
    journal = "Phys. Rev. D",
    volume = "74",
    pages = "123503",
    year = "2006"
}

@article{Carroll:2003st,
    author = "Carroll, Sean M. and Hoffman, Mark and Trodden, Mark",
    title = "{Can the dark energy equation-of-state parameter $w$ be less than $−1$?}",
    eprint = "astro-ph/0301273",
    archivePrefix = "arXiv",
    reportNumber = "EFI-2003-01, SU-GP-03-1-1",
    doi = "10.1103/PhysRevD.68.023509",
    journal = "Phys. Rev. D",
    volume = "68",
    pages = "023509",
    year = "2003"
}

@article{DESI:2025zgx,
    author = "Abdul Karim, M. and others",
    collaboration = "DESI",
    title = "{DESI DR2 results. II. Measurements of baryon acoustic oscillations and cosmological constraints}",
    eprint = "2503.14738",
    archivePrefix = "arXiv",
    primaryClass = "astro-ph.CO",
    reportNumber = "FERMILAB-PUB-25-0169-PPD",
    doi = "10.1103/tr6y-kpc6",
    journal = "Phys. Rev. D",
    volume = "112",
    number = "8",
    pages = "083515",
    year = "2025"
}

@ARTICLE{2022ScPA....2....1H,
       author = {{Haude}, Sophia and {Salehi}, Shabnam and {Vidal}, Sof{\'\i}a and {Maturi}, Matteo and {Bartelmann}, Matthias},
        title = "{Model-independent determination of the cosmic growth factor}",
      journal = {SciPost Astronomy},
         year = 2022,
        month = jan,
       volume = {2},
       number = {1},
          eid = {001},
        pages = {001},
          doi = {10.21468/SciPostAstro.2.1.001},
archivePrefix = {arXiv},
       eprint = {1912.04560},
 primaryClass = {astro-ph.CO},
       adsurl = {https://ui.adsabs.harvard.edu/abs/2022ScPA....2....1H}
}

@book{Ryden:1970vsj,
    author = "Ryden, B.",
    title = "{Introduction to cosmology}",
    doi = "10.1017/9781316651087",
    isbn = "978-1-107-15483-4, 978-1-316-88984-8, 978-1-316-65108-7",
    publisher = "Cambridge University Press",
    year = "2006"
}

@article{Jones:2009yz,
    author = "Jones, D. Heath and others",
    title = "{The 6dF Galaxy Survey: Final Redshift Release (DR3) and Southern Large-Scale Structures}",
    eprint = "0903.5451",
    archivePrefix = "arXiv",
    primaryClass = "astro-ph.CO",
    doi = "10.1111/j.1365-2966.2009.15338.x",
    journal = "Mon. Not. Roy. Astron. Soc.",
    volume = "399",
    pages = "683",
    year = "2009"
}

@article{Howlett:2017asq,
    author = {Howlett, Cullan and Staveley-Smith, Lister and Elahi, Pascal J. and Hong, Tao and Jarrett, Tom H. and Jones, D. Heath and Koribalski, B{\"a}rbel S. and Macri, Lucas M. and Masters, Karen L. and Springob, Christopher M.},
    title = "{2MTF {\textendash} VI. Measuring the velocity power spectrum}",
    eprint = "1706.05130",
    archivePrefix = "arXiv",
    primaryClass = "astro-ph.CO",
    doi = "10.1093/mnras/stx1521",
    journal = "Mon. Not. Roy. Astron. Soc.",
    volume = "471",
    number = "3",
    pages = "3135--3151",
    year = "2017"
}

@article{Beutler:2011hx,
    author = "Beutler, Florian and Blake, Chris and Colless, Matthew and Jones, D. Heath and Staveley-Smith, Lister and Campbell, Lachlan and Parker, Quentin and Saunders, Will and Watson, Fred",
    title = "{The 6dF Galaxy Survey: Baryon Acoustic Oscillations and the Local Hubble Constant}",
    eprint = "1106.3366",
    archivePrefix = "arXiv",
    primaryClass = "astro-ph.CO",
    doi = "10.1111/j.1365-2966.2011.19250.x",
    journal = "Mon. Not. Roy. Astron. Soc.",
    volume = "416",
    pages = "3017--3032",
    year = "2011"
}

@article{Ross:2014qpa,
    author = "Ross, Ashley J. and Samushia, Lado and Howlett, Cullan and Percival, Will J. and Burden, Angela and Manera, Marc",
    title = "{The clustering of the SDSS DR7 main Galaxy sample \textendash{} I. A 4 per cent distance measure at $z = 0.15$}",
    eprint = "1409.3242",
    archivePrefix = "arXiv",
    primaryClass = "astro-ph.CO",
    doi = "10.1093/mnras/stv154",
    journal = "Mon. Not. Roy. Astron. Soc.",
    volume = "449",
    number = "1",
    pages = "835--847",
    year = "2015"
}

@article{Ishak:2005zs,
    author = "Ishak, Mustapha and Upadhye, Amol and Spergel, David N.",
    title = "{Probing cosmic acceleration beyond the equation of state: Distinguishing between dark energy and modified gravity models}",
    eprint = "astro-ph/0507184",
    archivePrefix = "arXiv",
    doi = "10.1103/PhysRevD.74.043513",
    journal = "Phys. Rev. D",
    volume = "74",
    pages = "043513",
    year = "2006"
}

@article{Dvali:2000hr,
    author = "Dvali, G. R. and Gabadadze, Gregory and Porrati, Massimo",
    title = "{4-D gravity on a brane in 5-D Minkowski space}",
    eprint = "hep-th/0005016",
    archivePrefix = "arXiv",
    reportNumber = "NYU-TH-00-04-01",
    doi = "10.1016/S0370-2693(00)00669-9",
    journal = "Phys. Lett. B",
    volume = "485",
    pages = "208--214",
    year = "2000"
}

@article{Planck:2019nip,
    author = "Aghanim, N. and others",
    collaboration = "Planck",
    title = "{Planck 2018 results. V. CMB power spectra and likelihoods}",
    eprint = "1907.12875",
    archivePrefix = "arXiv",
    primaryClass = "astro-ph.CO",
    doi = "10.1051/0004-6361/201936386",
    journal = "Astron. Astrophys.",
    volume = "641",
    pages = "A5",
    year = "2020"
}

@article{Riess:2021jrx,
    author = "Riess, Adam G. and others",
    title = "{A Comprehensive Measurement of the Local Value of the Hubble Constant with 1 km s$^{-1}$ Mpc$^{-1}$ Uncertainty from the Hubble Space Telescope and the SH0ES Team}",
    eprint = "2112.04510",
    archivePrefix = "arXiv",
    primaryClass = "astro-ph.CO",
    doi = "10.3847/2041-8213/ac5c5b",
    journal = "Astrophys. J. Lett.",
    volume = "934",
    number = "1",
    pages = "L7",
    year = "2022"
}

@article{DES:2026fyc,
    author = "Abbott, T. M. C. and others",
    collaboration = "DES",
    title = "{Dark Energy Survey Year 6 Results: Cosmological Constraints from Galaxy Clustering and Weak Lensing}",
    eprint = "2601.14559",
    archivePrefix = "arXiv",
    primaryClass = "astro-ph.CO",
    reportNumber = "DES-2025-0929, FERMILAB-PUB-26-0026-PPD",
    month = "1",
    year = "2026"
}

@article{Pisanti:2007hk,
    author = "Pisanti, O. and Cirillo, A. and Esposito, S. and Iocco, F. and Mangano, G. and Miele, G. and Serpico, P. D.",
    title = "{PArthENoPE: Public Algorithm Evaluating the Nucleosynthesis of Primordial Elements}",
    eprint = "0705.0290",
    archivePrefix = "arXiv",
    primaryClass = "astro-ph",
    reportNumber = "DSF-13-07, FERMILAB-PUB-07-079-A, SLAC-PUB-12488",
    doi = "10.1016/j.cpc.2008.02.015",
    journal = "Comput. Phys. Commun.",
    volume = "178",
    pages = "956--971",
    year = "2008"
}

@article{Taruya:2010mx,
    author = "Taruya, Atsushi and Nishimichi, Takahiro and Saito, Shun",
    title = "{Baryon Acoustic Oscillations in 2D: Modeling Redshift-space Power Spectrum from Perturbation Theory}",
    eprint = "1006.0699",
    archivePrefix = "arXiv",
    primaryClass = "astro-ph.CO",
    doi = "10.1103/PhysRevD.82.063522",
    journal = "Phys. Rev. D",
    volume = "82",
    pages = "063522",
    year = "2010"
}

@article{Bartelmann:1999yn,
    author = "Bartelmann, Matthias and Schneider, Peter",
    title = "{Weak gravitational lensing}",
    eprint = "astro-ph/9912508",
    archivePrefix = "arXiv",
    doi = "10.1016/S0370-1573(00)00082-X",
    journal = "Phys. Rept.",
    volume = "340",
    pages = "291--472",
    year = "2001"
}

@article{Perenon:2022fgw,
    author = "Perenon, Louis and Martinelli, Matteo and Maartens, Roy and Camera, Stefano and Clarkson, Chris",
    title = "{Measuring dark energy with expansion and growth}",
    eprint = "2206.12375",
    archivePrefix = "arXiv",
    primaryClass = "astro-ph.CO",
    doi = "10.1016/j.dark.2022.101119",
    journal = "Phys. Dark Univ.",
    volume = "37",
    pages = "101119",
    year = "2022"
}

@article{DES:2020iqt,
    author = "Muir, J. and others",
    collaboration = "DES",
    title = "{DES Y1 results: Splitting growth and geometry to test $\Lambda$CDM}",
    eprint = "2010.05924",
    archivePrefix = "arXiv",
    primaryClass = "astro-ph.CO",
    reportNumber = "FERMILAB-PUB-20-529-AE, DES-2019-0520",
    doi = "10.1103/PhysRevD.103.023528",
    journal = "Phys. Rev. D",
    volume = "103",
    number = "2",
    pages = "023528",
    year = "2021"
}

@article{DES:2018ioi,
    author = "Brout, D. and others",
    collaboration = "DES",
    title = "{First Cosmology Results Using Type Ia Supernovae From the Dark Energy Survey: Analysis, Systematic Uncertainties, and Validation}",
    eprint = "1811.02377",
    archivePrefix = "arXiv",
    primaryClass = "astro-ph.CO",
    reportNumber = "FERMILAB-PUB-18-541-AE",
    doi = "10.3847/1538-4357/ab08a0",
    journal = "Astrophys. J.",
    volume = "874",
    pages = "150",
    year = "2019"
}

@article{PhysRevD.105.023520,
  title = {Dark Energy Survey Year 3 results: Cosmological constraints from galaxy clustering and weak lensing},
  author = {Abbott, T. M. C. and Aguena, M. and Alarcon, A. and Allam, S. and Alves, O. and Amon, A. and Andrade-Oliveira, F. and Annis, J. and Avila, S. and Bacon, D. and Baxter, E. and Bechtol, K. and Becker, M. R. and Bernstein, G. M. and Bhargava, S. and Birrer, S. and Blazek, J. and Brandao-Souza, A. and Bridle, S. L. and Brooks, D. and Buckley-Geer, E. and Burke, D. L. and Camacho, H. and Campos, A. and Carnero Rosell, A. and Carrasco Kind, M. and Carretero, J. and Castander, F. J. and Cawthon, R. and Chang, C. and Chen, A. and Chen, R. and Choi, A. and Conselice, C. and Cordero, J. and Costanzi, M. and Crocce, M. and da Costa, L. N. and da Silva Pereira, M. E. and Davis, C. and Davis, T. M. and De Vicente, J. and DeRose, J. and Desai, S. and Di Valentino, E. and Diehl, H. T. and Dietrich, J. P. and Dodelson, S. and Doel, P. and Doux, C. and Drlica-Wagner, A. and Eckert, K. and Eifler, T. F. and Elsner, F. and Elvin-Poole, J. and Everett, S. and Evrard, A. E. and Fang, X. and Farahi, A. and Fernandez, E. and Ferrero, I. and Fert\'e, A. and Fosalba, P. and Friedrich, O. and Frieman, J. and Garc\'{\i}a-Bellido, J. and Gatti, M. and Gaztanaga, E. and Gerdes, D. W. and Giannantonio, T. and Giannini, G. and Gruen, D. and Gruendl, R. A. and Gschwend, J. and Gutierrez, G. and Harrison, I. and Hartley, W. G. and Herner, K. and Hinton, S. R. and Hollowood, D. L. and Honscheid, K. and Hoyle, B. and Huff, E. M. and Huterer, D. and Jain, B. and James, D. J. and Jarvis, M. and Jeffrey, N. and Jeltema, T. and Kovacs, A. and Krause, E. and Kron, R. and Kuehn, K. and Kuropatkin, N. and Lahav, O. and Leget, P.-F. and Lemos, P. and Liddle, A. R. and Lidman, C. and Lima, M. and Lin, H. and MacCrann, N. and Maia, M. A. G. and Marshall, J. L. and Martini, P. and McCullough, J. and Melchior, P. and Mena-Fern\'andez, J. and Menanteau, F. and Miquel, R. and Mohr, J. J. and Morgan, R. and Muir, J. and Myles, J. and Nadathur, S. and Navarro-Alsina, A. and Nichol, R. C. and Ogando, R. L. C. and Omori, Y. and Palmese, A. and Pandey, S. and Park, Y. and Paz-Chinch\'on, F. and Petravick, D. and Pieres, A. and Plazas Malag\'on, A. A. and Porredon, A. and Prat, J. and Raveri, M. and Rodriguez-Monroy, M. and Rollins, R. P. and Romer, A. K. and Roodman, A. and Rosenfeld, R. and Ross, A. J. and Rykoff, E. S. and Samuroff, S. and S\'anchez, C. and Sanchez, E. and Sanchez, J. and Sanchez Cid, D. and Scarpine, V. and Schubnell, M. and Scolnic, D. and Secco, L. F. and Serrano, S. and Sevilla-Noarbe, I. and Sheldon, E. and Shin, T. and Smith, M. and Soares-Santos, M. and Suchyta, E. and Swanson, M. E. C. and Tabbutt, M. and Tarle, G. and Thomas, D. and To, C. and Troja, A. and Troxel, M. A. and Tucker, D. L. and Tutusaus, I. and Varga, T. N. and Walker, A. R. and Weaverdyck, N. and Wechsler, R. and Weller, J. and Yanny, B. and Yin, B. and Zhang, Y. and Zuntz, J.},
  collaboration = {DES Collaboration},
  journal = {Phys. Rev. D},
  volume = {105},
  issue = {2},
  pages = {023520},
  numpages = {42},
  year = {2022},
  month = {Jan},
  publisher = {American Physical Society},
  doi = {10.1103/PhysRevD.105.023520},
  url = {https://link.aps.org/doi/10.1103/PhysRevD.105.023520}
}

@article{DES:2024jxu,
    author = "Abbott, T. M. C. and others",
    collaboration = "DES",
    title = "{The Dark Energy Survey: Cosmology Results with \ensuremath{\sim}1500 New High-redshift Type Ia Supernovae Using the Full 5 yr Data Set}",
    eprint = "2401.02929",
    archivePrefix = "arXiv",
    primaryClass = "astro-ph.CO",
    reportNumber = "FERMILAB-PUB-23-0821-PPD, DES-2023-805",
    doi = "10.3847/2041-8213/ad6f9f",
    journal = "Astrophys. J. Lett.",
    volume = "973",
    number = "1",
    pages = "L14",
    year = "2024"
}

@article{Fang:2019xat,
    author = "Fang, Xiao and Krause, Elisabeth and Eifler, Tim and MacCrann, Niall",
    title = "{Beyond Limber: Efficient computation of angular power spectra for galaxy clustering and weak lensing}",
    eprint = "1911.11947",
    archivePrefix = "arXiv",
    primaryClass = "astro-ph.CO",
    doi = "10.1088/1475-7516/2020/05/010",
    journal = "JCAP",
    volume = "05",
    pages = "010",
    year = "2020"
}

@article{Planck:2018vyg,
    author = "Aghanim, N. and others",
    collaboration = "Planck",
    title = "{Planck 2018 results. VI. Cosmological parameters}",
    eprint = "1807.06209",
    archivePrefix = "arXiv",
    primaryClass = "astro-ph.CO",
    doi = "10.1051/0004-6361/201833910",
    journal = "Astron. Astrophys.",
    volume = "641",
    pages = "A6",
    year = "2020",
    note = "[Erratum: Astron.Astrophys. 652, C4 (2021)]"
}

@article{Blanchard:2022xkk,
    author = "Blanchard, Alain and H{\'e}loret, Jean-Yves and Ili{\'c}, St{\'e}phane and Lamine, Brahim and Tutusaus, Isaac",
    title = "{$\Lambda$CDM is alive and well}",
    eprint = "2205.05017",
    archivePrefix = "arXiv",
    primaryClass = "astro-ph.CO",
    doi = "10.33232/001c.117170",
    journal = "Open J. Astrophys.",
    volume = "7",
    pages = "117170",
    year = "2024"
}

@article{Howlett:2012mh,
    author = "Howlett, Cullan and Lewis, Antony and Hall, Alex and Challinor, Anthony",
    title = "{CMB power spectrum parameter degeneracies in the era of precision cosmology}",
    eprint = "1201.3654",
    archivePrefix = "arXiv",
    primaryClass = "astro-ph.CO",
    doi = "10.1088/1475-7516/2012/04/027",
    journal = "JCAP",
    volume = "04",
    pages = "027",
    year = "2012"
}

@article{MacCrann:2019ntb,
    author = "MacCrann, Niall and Blazek, Jonathan and Jain, Bhuvnesh and Krause, Elisabeth",
    title = "{Controlling and leveraging small-scale information in tomographic galaxy{\textendash}galaxy lensing}",
    eprint = "1903.07101",
    archivePrefix = "arXiv",
    primaryClass = "astro-ph.CO",
    doi = "10.1093/mnras/stz2761",
    journal = "Mon. Not. Roy. Astron. Soc.",
    volume = "491",
    number = "4",
    pages = "5498--5509",
    year = "2020"
}

@article{McEwen:2016fjn,
    author = "McEwen, Joseph E. and Fang, Xiao and Hirata, Christopher M. and Blazek, Jonathan A.",
    title = "{FAST-PT: a novel algorithm to calculate convolution integrals in cosmological perturbation theory}",
    eprint = "1603.04826",
    archivePrefix = "arXiv",
    primaryClass = "astro-ph.CO",
    doi = "10.1088/1475-7516/2016/09/015",
    journal = "JCAP",
    volume = "09",
    pages = "015",
    year = "2016"
}

@article{Fang:2016wcf,
    author = "Fang, Xiao and Blazek, Jonathan A. and McEwen, Joseph E. and Hirata, Christopher M.",
    title = "{FAST-PT II: an algorithm to calculate convolution integrals of general tensor quantities in cosmological perturbation theory}",
    eprint = "1609.05978",
    archivePrefix = "arXiv",
    primaryClass = "astro-ph.CO",
    doi = "10.1088/1475-7516/2017/02/030",
    journal = "JCAP",
    volume = "02",
    pages = "030",
    year = "2017"
}

@article{Hirata:2004gc,
    author = "Hirata, Christopher M. and Seljak, Uros",
    title = "{Intrinsic alignment-lensing interference as a contaminant of cosmic shear}",
    eprint = "astro-ph/0406275",
    archivePrefix = "arXiv",
    doi = "10.1103/PhysRevD.82.049901",
    journal = "Phys. Rev. D",
    volume = "70",
    pages = "063526",
    year = "2004",
    note = "[Erratum: Phys.Rev.D 82, 049901 (2010)]"
}

@article{Bridle:2007ft,
    author = "Bridle, Sarah and King, Lindsay",
    title = "{Dark energy constraints from cosmic shear power spectra: impact of intrinsic alignments on photometric redshift requirements}",
    eprint = "0705.0166",
    archivePrefix = "arXiv",
    primaryClass = "astro-ph",
    doi = "10.1088/1367-2630/9/12/444",
    journal = "New J. Phys.",
    volume = "9",
    pages = "444",
    year = "2007"
}

@article{Lewis:1999bs,
    author = "Lewis, Antony and Challinor, Anthony and Lasenby, Anthony",
    title = "{Efficient computation of CMB anisotropies in closed FRW models}",
    eprint = "astro-ph/9911177",
    archivePrefix = "arXiv",
    doi = "10.1086/309179",
    journal = "Astrophys. J.",
    volume = "538",
    pages = "473--476",
    year = "2000"
}

@article{Zuntz:2014csq,
    author = "Zuntz, Joe and Paterno, Marc and Jennings, Elise and Rudd, Douglas and Manzotti, Alessandro and Dodelson, Scott and Bridle, Sarah and Sehrish, Saba and Kowalkowski, James",
    title = "{CosmoSIS: modular cosmological parameter estimation}",
    eprint = "1409.3409",
    archivePrefix = "arXiv",
    primaryClass = "astro-ph.CO",
    reportNumber = "FERMILAB-PUB-14-408-A",
    doi = "10.1016/j.ascom.2015.05.005",
    journal = "Astron. Comput.",
    volume = "12",
    pages = "45--59",
    year = "2015"
}

@article{Ruiz:2014hma,
    author = "Ruiz, Eduardo J. and Huterer, Dragan",
    title = "{Testing the dark energy consistency with geometry and growth}",
    eprint = "1410.5832",
    archivePrefix = "arXiv",
    primaryClass = "astro-ph.CO",
    doi = "10.1103/PhysRevD.91.063009",
    journal = "Phys. Rev. D",
    volume = "91",
    pages = "063009",
    year = "2015"
}

@article{Lewis:2019xzd,
    author = "Lewis, Antony",
    title = "{GetDist: a Python package for analysing Monte Carlo samples}",
    eprint = "1910.13970",
    archivePrefix = "arXiv",
    primaryClass = "astro-ph.IM",
    doi = "10.1088/1475-7516/2025/08/025",
    journal = "JCAP",
    volume = "08",
    pages = "025",
    year = "2025"
}

@article{Colgain:2022rxy,
    author = "Colg{\'a}in, Eoin {\'O}. and Sheikh-Jabbari, M. M. and Solomon, Rance and Dainotti, Maria G. and Stojkovic, Dejan",
    title = "{Putting flat {\ensuremath{\Lambda}}CDM in the (Redshift) bin}",
    eprint = "2206.11447",
    archivePrefix = "arXiv",
    primaryClass = "astro-ph.CO",
    doi = "10.1016/j.dark.2024.101464",
    journal = "Phys. Dark Univ.",
    volume = "44",
    pages = "101464",
    year = "2024"
}

@article{Zhong:2023how,
    author = "Zhong, Kunhao and Saraivanov, Evan and Miranda, Vivian and Xu, Jiachuan and Eifler, Tim and Krause, Elisabeth",
    title = "{Growth and geometry split in light of the DES-Y3 survey}",
    eprint = "2301.03694",
    archivePrefix = "arXiv",
    primaryClass = "astro-ph.CO",
    doi = "10.1103/PhysRevD.107.123529",
    journal = "Phys. Rev. D",
    volume = "107",
    number = "12",
    pages = "123529",
    year = "2023"
}

@article{Ruiz-Zapatero:2021rzl,
    author = "Ruiz-Zapatero, Jaime and others",
    title = "{Geometry versus growth - Internal consistency of the flat $\Lambda$CDM model with KiDS-1000}",
    eprint = "2105.09545",
    archivePrefix = "arXiv",
    primaryClass = "astro-ph.CO",
    doi = "10.1051/0004-6361/202141350",
    journal = "Astron. Astrophys.",
    volume = "655",
    pages = "A11",
    year = "2021"
}

@article{Campos:2022gtg,
    author = "Campos, Andresa and Samuroff, Simon and Mandelbaum, Rachel",
    title = "{An empirical approach to model selection: weak lensing and intrinsic alignments}",
    eprint = "2211.02800",
    archivePrefix = "arXiv",
    primaryClass = "astro-ph.CO",
    doi = "10.1093/mnras/stad2213",
    journal = "Mon. Not. Roy. Astron. Soc.",
    volume = "525",
    number = "2",
    pages = "1885--1901",
    year = "2023"
}
\bibliographystyle{aa}

\begin{appendix}
\section{Galaxy-galaxy lensing}
\label{appendix-ggl}
\begin{figure}[h]
	\centering
	\includegraphics[width=\linewidth]{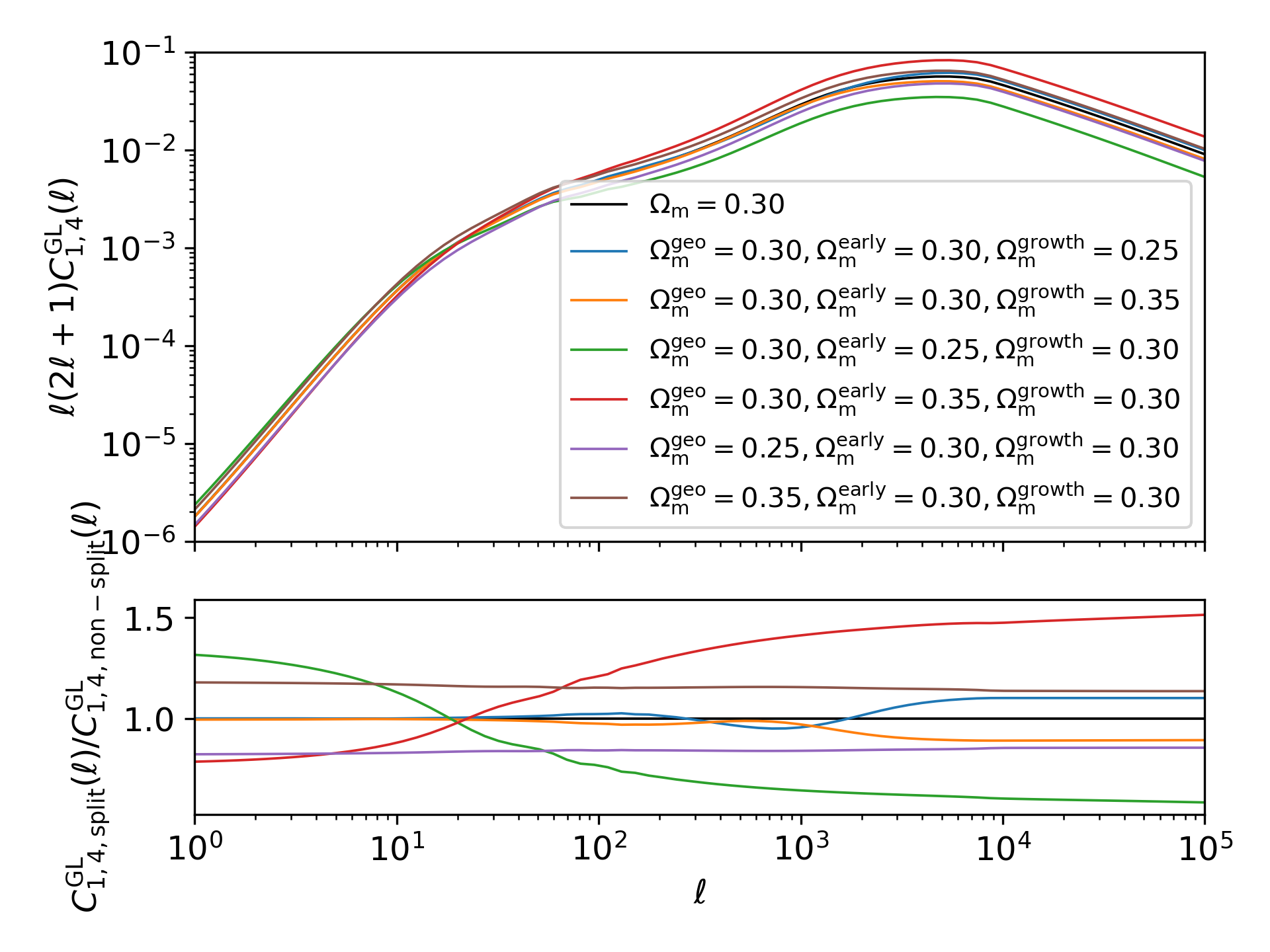}
    \caption{Top: Split angular power spectrum of GGL for the cross-correlation of the first lens bin and the last DES Y3 source bin (1,4) with different values for $\Om$ in the three regimes, showing the dependence on geometry, growth, and the early Universe. Bottom: Ratio of the split angular power spectrum of WL with respect to the non-split angular power spectrum at $\Om=0.3$.}
	\label{fig:C_ell_GGL}
\end{figure}

The GGL angular power spectrum is influenced by all three regimes, as we show in Fig.\,\ref{fig:C_ell_GGL}.

\begin{figure}
	\centering
	\includegraphics[width=\linewidth]{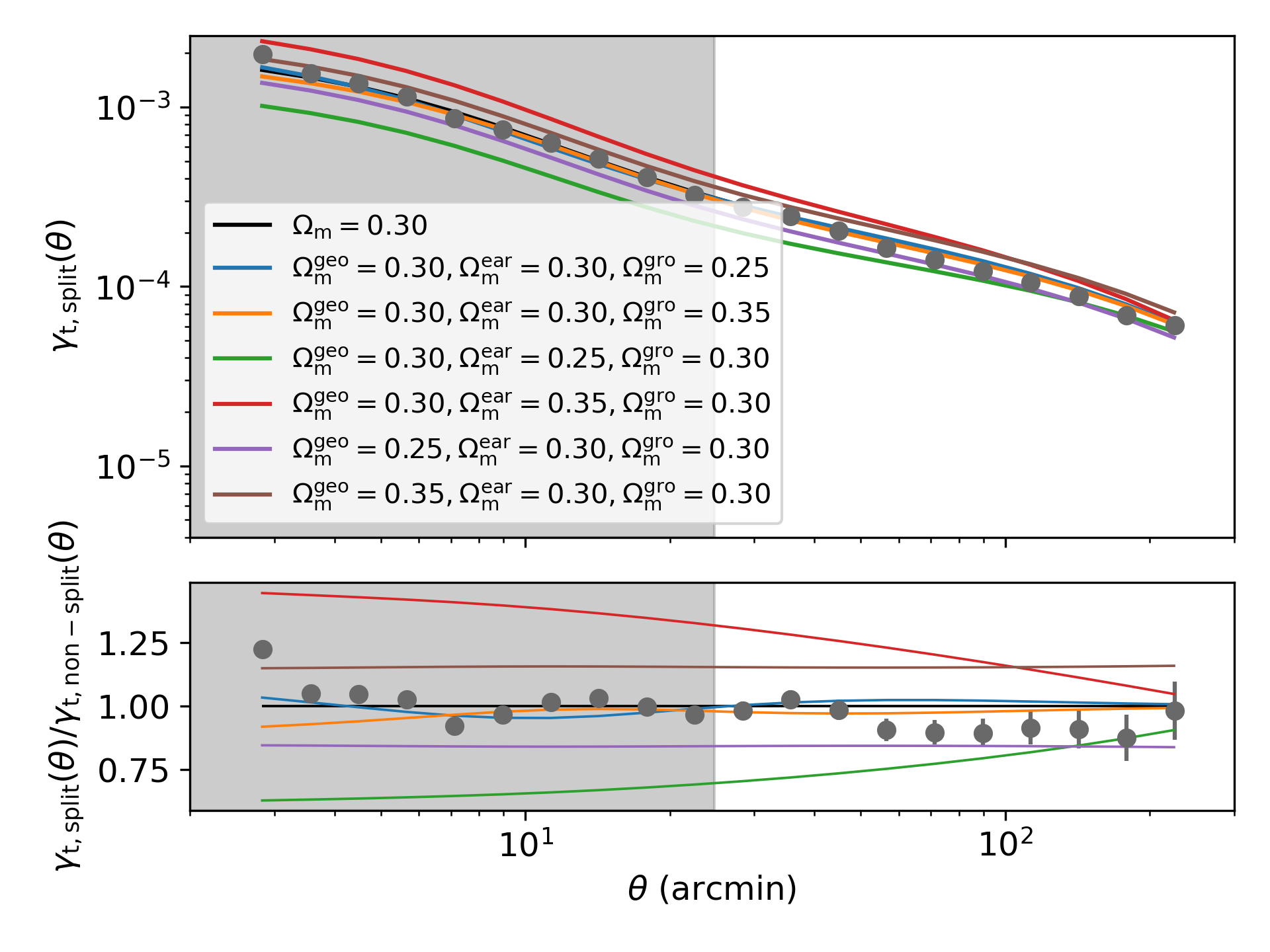}
	\caption{Top: Split two-point correlation function for GGL with DES Y3 data in the tomographic bin (1,4), that is, the lenses in bin 1 and the sources in bin 4, including theoretical predictions with different $\Om$ values per regime. The DES Y3 data points are shown in grey and the grey band indicates the scale cuts. Bottom: Ratio of the split correlation function with respect to the non-split correlation function with $\Omega_{\rm m}=0.3$.}
	\label{fig:gamma-t}
\end{figure}

This naturally resembles the GC and WL angular power spectra from Figs. \ref{fig:C_ell_GC} and \ref{fig:C_ell_WL}. We show that the early regime (in green and red) has the biggest influence through the matter power spectrum. The geometry regime (in purple and brown) also has an influence that is less strong than for the WL case. This is because $\Om^{\rm geo}$ entered only in one of the window functions.

We show the prediction and data points of the two-point correlation function for GGL using the bin combination with the highest S/N. This is the one where the sources were in the farthest bin, here bin 4 ($0.87<z<2.0$), and the lenses in the closest bin, bin 1 ($0.15<z<0.35$). This is shown in Fig.\,\ref{fig:gamma-t}, where we show the theoretical prediction varying $\Om$ in each regime and compare them to the DES Y3 measurements. Similarly to the GC and WL data, this shows a very small influence of the growth regime, a tilt when varying the early regime, and a shift when varying the geometric regime.

\section{Weak-lensing window function}
\label{appendix-wl-window}

The window function for WL, see Eq.\,(\ref{wl-window}), contains the $\Om$ parameter. To see whether this falls into the geometry or the growth regime, we considered its derivation from the effective convergence. This follows \cite{Bartelmann:1999yn}. The effective convergence in the thin-lens approximation and a flat cosmology is:

\begin{equation}
\kappa_{\rm eff}(\vec{\theta}, \chi) = \frac{1}{c^2}\int_0^\chi d\chi' \frac{\chi' (\chi-\chi')}{\chi} \Delta_r (\delta \Phi)\,.
\label{eff_conv}
\end{equation}

Here, $\vec{\theta}$ is the sky direction in which the light rays propagate, $\Delta_r$ is the Laplacian and $\delta\Phi$ is the potential difference between two light rays coming from the Born approximation. We used the Poisson equation with the Newtonian potential $\Phi$, the gravitational constant $G_{\rm N}$, and the density $\rho$:

\begin{equation}
    \Delta_r \Phi = 4\pi G_{\rm N} \rho\,.
\end{equation}

This can be split into a background and a perturbation potential:
\begin{equation}
    \Delta_r (\bar{\Phi} + \delta\Phi) = 4\pi G_{\rm N}(\bar{\rho}+\delta \rho)\, ,
\end{equation}
with the background density $\bar{\rho}$ and the density perturbations $\delta \rho$. We considered the perturbation part since this concerns the difference between two light rays and transformed the Laplacian to comoving coordinates:
\begin{equation}
     \Delta_\chi (\delta\Phi )= 4\pi G_{\rm N}\bar{\rho} \delta_{\rm m} a^2\,.
     \label{perturbation-poisson-comoving}
\end{equation}

At this point, we wrote the background density in terms of $\Om$ in the matter-dominated era. We replaced the background density using the geometric matter density:
\begin{equation}
    \bar{\rho}=a^{-3}\rho_{\rm crit}\Omega_{\rm m}^{\rm geo}\,.
\end{equation}

This can be inserted into Eq.\,(\ref{perturbation-poisson-comoving}):
\begin{equation}
    \Delta_\chi(\delta \Phi) = 4\pi G_{\rm N} \Omega_{\rm m}^{\rm geo} \rho_{\rm crit} \delta_{\rm m} a^{-1}\,.
    \label{poisson-delta-m}
\end{equation}

The critical density can also be rewritten as $\rho_{\rm crit}=3 H_0^2/(8\pi G_{\rm N})$. Inserting this Laplacian into Eq.\,(\ref{eff_conv}), we obtained:
\begin{equation}
     \kappa_{\rm eff}(\vec{\theta}, \chi) = \frac{3 H_0^2 \Omega_{\rm m}^{\rm geo}}{2 c^2} \int_0^\chi d\chi' \frac{\chi' (\chi-\chi')}{\chi} \frac{\delta_{\rm m}(\chi' \vec{\theta},\chi')}{a(\chi')}\,.
\end{equation}

This led to using the geometric regime, that is, the $\Omega_m^{\rm geo}$ parameter, in the window function when computing WL angular power spectra.

\begin{figure*}[b]
    \centering
    \sidecaption
	\includegraphics[width=12cm]{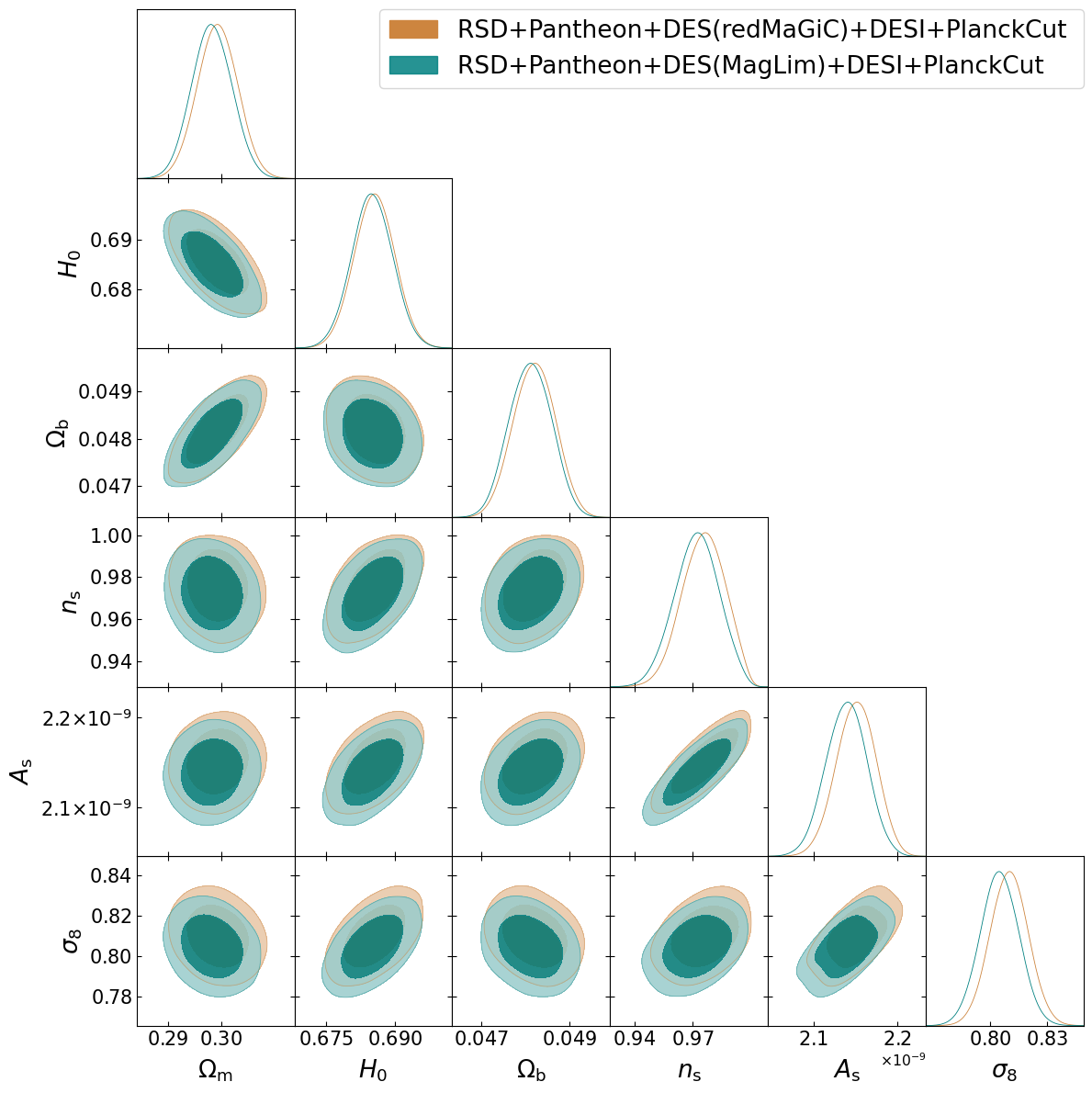}
	\caption{Comparison of the cosmological constraints for 3x2pt with the different DES Y3 lens samples, MagLim in teal and redMaGiC in ochre, combined with all other probes, that is, \textit{Planck} CMB with scale cuts, DESI DR2 BAO, Pantheon+ SNe Ia, and RSD.}
	\label{fig:lens-sample}
\end{figure*}

\section{Nuisance Parameters}
\subsection{Dark Energy Survey Y3}
\label{nuisance-des}

As in the DES Y3 analysis, we varied a number of nuisance parameters that capture astrophysical and systematic effects on the measurements. Each lens galaxy bin has a linear galaxy bias parameter $b_i$. For IA, we used NLA, as described above, which contains two additional nuisance parameters. All lens and source redshift bins contain an additive shift to the mean redshift $\Delta z_i$. This results in the redshift distribution $n_i(z)$:
\begin{equation}
    n_i(z)=n^{\rm photo-z}_i(z-\Delta z_i).
\end{equation}

Here, $n_i^{\rm photo-z}$ refers to the photometric redshift distribution. Lastly, the shear calibration bias added another nuisance parameter per bin. This is because the measured ellipticity $e^j_i$ depends on the true shear $\gamma^j_i$ through the multiplicative bias $m_i$:

\begin{equation}
    e^j_i=(1+m_i)\gamma^j_i\,.
\end{equation}

\subsection{\textit{Planck} 2018}
\label{nuisance-planck}
We further varied the same nuisance parameters as in the \textit{Planck} analysis \citep{Planck:2019nip}. Firstly, we included the point-source contribution in different frequency bands (100GHz$\times$100GHz, 143GHz$\times$143GHz, 143GHz$\times$217GHz, and 217GHz$\times$217GHz). In the same frequency combinations at $\ell=200$, we varied the dust residual contamination for the TT spectra. For the TE spectra, the dust residual contamination was varied for all six combinations of the three frequencies (100GHz, 143GHz, and 217 GHz). The relative calibration for the temperature spectrum posed two additional nuisance parameters, one between 100 and 143 GHz and one between 143 and 217 GHz. We also varied the Cosmic Infrared Background contamination, the thermal Sunyaev–Zeldovich effect and their cross-correlation. The last nuisance parameter was the absolute \textit{Planck} calibration.

\section{DES Y3 lens sample}
\label{appendix-maglim}

We used the redMaGiC lens sample as our baseline for the DES Y3 data. However, both the redMaGiC and MagLim samples were considered in the DES Y3 analysis \citep{PhysRevD.105.023520}. In Fig.\,\ref{fig:lens-sample}, we compared the two samples when using all observables (DES Y3, \textit{Planck} with scale cuts, DESI DR2, Pantheon+, and RSD). All contours overlap and the difference between the two is minimal.

\end{appendix}
\end{document}